\documentclass[aps,prb,superscriptaddress,showpacs,floatfix]{revtex4-1}
\usepackage{graphicx,amsmath,amssymb,xspace,epsfig,float,multirow,subfigure,tabularx}
\usepackage{amsbsy,amssymb,amsmath,bm}
\usepackage{epsf}
\usepackage{color,natbib}
\usepackage{amsfonts}
\usepackage{hyperref}

\pdfoutput=1

\begin{document}

\title{Apical oxygen vibrations dominant role in d-wave cuprate
superconductivity and its interplay with spin fluctuations.}
\author{B. Rosenstein}

\affiliation{Department of Electrohysics, National Yang Ming Chiao Tung University, Hsinchu,
Taiwan, R.O.C. }

\author{B. Ya. Shapiro }

\affiliation{Department of Physics, Institute of Superconductivity, Bar-Ilan
University, 52900 Ramat-Gan, Israel.}

\begin{abstract}
Microscopic theory of a high $T_{c}$ cuprate $Bi_{2}Sr_{2}CaCu_{2}O_{8+x}$
based on main pairing channel\ of electrons in $CuO$ planes due to $40mev$
lateral vibrations of the apical oxygen atoms in adjacent the $SrO$ ionic
insulator layer is proposed. The separation between the vibrating charged
atoms and the 2D electron gas creates the forward scattering peak leading in
turn to the d -wave pairing within Eliashberg formalism. The phonon mode
naturally explain the kink in dispersion relation observed by ARPES and the
and effect of the $O^{16}\rightarrow O^{18}$ isotope substitution in the
normal state. To describe the pseudogap physics a single band fourfold
symmetric $t-t^{\prime }$ Hubbard model, with the hopping parameters $%
t^{\prime }\sim -0.17t$ and the on site repulsion e $U\sim 6t$. It described
the $\,$Mott insulator at low doping, while at higher dopping the pseudogap
physics (still strongly correlated)\ can be be approximated by the
symmetrized mean field model and with renormalized $U$ incorporating
screening. The location of the transition line $T^{\ast }$between the
locally antiferromagnetic pseudogap and the paramagnetic overdoped phases
and susceptibility (describing spin fluctuations coupling to 2DEG) are also
obtained within this approximation. The superconducting $d$ - wave gap
mainly due to the phonon channel but is assisted by the spin fluctuations
(15-20\%). The dependence of the gap and $T_{c}$ on doping and effect of the
\ isotope substitution are obtained and is consistent with experiments.
\end{abstract}

\maketitle

\section{Introduction.}

\textit{\ }For decades the only\textit{\ }superconductors with critical
temperature above $90K$ under ambient conditions were cuprates like $%
Bi_{2}Sr_{2}CaCu_{2}O_{8+x}$ ($Bi2212$). They are generally characterized by
the following five structural/chemical/electronic peculiarities. First, they
are all quasi - two dimensional (2D) perovskite layered oxides. Second, the
2D electron gas (2DEG) in which the superconductivity resides is created by
"charging" $CuO$ planes: hole doping the anti - ferromagnetic (AF) parent
material. Third, the conducting layers are separated by several insulating
ionic oxide planes. Fourth, as doping decreases past optimal the pseudogap
is opened and closed Fermi surface splits into four arcs\cite{pseudogap} (a
topological transition). Fifth is the d - wave symmetry of the order
parameter below the "superconducting dome" on the phase diagram. It is
widely believed\cite{Dagotto1} that, although the insulating layers play a
role in charging the $CuO$ planes, the (still not clearly identified) bosons
responsible for the pairing (so called "glue") are confined to the $CuO$
layer.

Several years ago another group of superconducting materials with critical
temperature as high as $T_{c}=60-106K$ was fabricated by deposition of a
single unit cell layer (1UC) of $FeSe$ on insulating substrates like $%
SrTiO_{3}$ (STO both\cite{expFeSe} $\left( 001\right) $ and\cite{110} $%
\left( 110\right) $), $TiO_{2}$ and\cite{rutileFeSe} $BaTiO_{3}$. Note that
the first three of the characteristic cuprate features listed above are
manifest in these systems as well. Indeed, the insulating substrates are
again the perovskite oxide planes. The electron gas residing in the $FeSe$
layer\cite{charging} is charged (doped) by the perovskite substrate. The
remaining two of the five cuprate features are clearly distinct in the new
superconductor family. The Fermi surface is nearly round in sharp contrast
to the rhomb - shaped one in cuprates. There are neither pseudogap nor the
electron "pockets". Furthermore the symmetry of the order parameter is the
noddles s - wave\cite{swave}. Generally the system is much simpler than the
cuprates and much progress in understanding of its superconductivity
mechanism was achieved. The role of the insulating substrate in $FeSe/STO$
seems to extend beyond the charging \cite{charging}. While the physical
nature of the pairing boson in cuprates is still under discussion, it became
clear that superconductivity mechanism in 1UC $FeSe/STO$ should at least
include the \textit{substrate} phonon exchange. Although there are theories
based on an unconventional boson exchange \textit{within} the pnictides
plane (perhaps spin fluctuations exchange\cite{Leerev}, as in pnictides
theories\cite{Dagotto2}), an alternative point of view was clearly formed%
\cite{Gorkov,JohnsonNJP16} based on idea that the pairing in the $FeSe$
plane is largely due to vibration of oxygen atoms in a substrate oxide layer
near the interface.

Historically a smoking gun for the relevance of the electron - phonons
interactions (EPI) to superconductivity has been the isotope effect. When
the isotope $^{16}O$ in surface layers of the $STO$ substrate was substituted%
\cite{isotopeGuo} by $^{18}O$, the gap at low temperature ($6K$) decreased
by about 10\%. Detailed measurements of the phonon spectrum via electron
energy loss spectroscopy \cite{Xue16phonon} demonstrated that the interface
phonons are very energetic (the "hard" longitudinal optical (LO) branch
appears at $\Omega _{h}=100mev$). The phonons couple to 2DEG with relatively
small coupling constant\cite{isotopeGuo} $\lambda \simeq 0.25$, deduced from
the intensity of the replica bands identified by ARPES \cite{Lee12}.
Importantly the interpretation of the replica bands was based on the forward
peak in the electron - phonon scattering (FSP). Initially this inspired an
idea that the surface phonons alone could provide a sufficiently strong
pairing\cite{JohnsonNJP16}. Since the BCS scenario, $T_{c}\approx \Omega
_{h}e^{-1/\lambda }$, is clearly out, one had to look for other ideas like
the extreme, delta like, FSP model\cite{Kulicrev},\cite{Kulichearly} for
which $T_{c}\approx \frac{\lambda }{2+3\lambda }\Omega _{h}$. This lead\cite%
{JohnsonNJP16} to sufficiently high $T_{c}$ for small $\lambda $.
Unfortunately the EPI parameters to achieve such a strong FSP in ionic
substrate are unrealistic. In a recent work\cite{Rosen19} we developed a
sufficiently precise microscopic model of phonons in adjacent insulating $%
TiO_{2}$ layer of the STO substrate and found an additional $\Omega
_{s}=50mev$ LO interface phonon. Since coupling of the $\Omega _{s}$ to the
electron gas in the $FeSe$ layer is practically the same as that of the hard 
$\Omega _{h}$ mode, it greatly enhances pairing. The momentum dependence of
the EPI matrix elements has an exponential FSP, $exp\left[ -2pd_{a}\right] $%
, where $d_{a}$ is the distance between the ionic layer and 2DEG. Calculated
coupling $\lambda $, critical temperature, replica band and other
characteristics of the superconducting state are consistent with
experiments. It demonstrated that the perovskite ionic layer phonons
constitute a sufficiently strong "glue" to mediate high $T_{c}$
superconductivity.

A question arises whether similar phononic pairing mechanism occurs in
cuprates. Of course there is a structural difference between the cuprates
and the 1UC $FeSe/STO$ in that the the bulk layered cuprates contain many $%
CuO$ planes, while there is a single $FeSe$ layer. The difference turns out
to be insignificant, since it was demonstrated\cite{accurate,Kim2UC} that
even two unit cells of optimally doped $Bi2212$\ sandwiched between
insulating materials exhibits practically as high $T_{c}$ as the bulk
material. Also recently a $CuO$ monolayer on top of $Bi2212$ film was
synthesized\cite{XueBSCCO} with surprisingly high the critical temperature
of $100K$. The pairing is of a noddles s-wave variety as in 1UC $FeSe/STO$
in striking contrast with $Bi2212$ and other hole doped cuprates. The s -
wave symmetry was explained by extremely strong charging\cite{XueBSCCO}\cite%
{1UCCuotheory}. In particular it was noticed that the Fermi surface becomes
nearly \textit{circular}\cite{1UCCuotheory} also in sharp contrast to the
rhombic shape of hole doped cuprates.

The idea that phonons are at least partially responsible for the d - wave
pairing has been contemplated over the years. In particular the $CuO$ layer
oxygen atoms breathing and buckling modes\cite{Bulut} and the apical oxygen $%
c$ axis vibrations\cite{apicz} \cite{apictheory}\cite{smokinggun} have been
considered. It is well established that phonons cause s - wave pairing in
low $T_{c}$ materials, d-wave pairing is possible when FSP is present. It
turns out that the nature of pairing for the FSP phonons depend on the shape
of the Fermi surface, assumed to be fourfold symmetric throughout this
paper. Our experience can be summarizes as follows. The pairing tends to be
d-wave a for rhomb - like Fermi surface and s-wave for a more circular one
like that of 1UC $FeSe/STO$ or $CuO/Bi2212$. Early work in this direction
was summarized in ref. \cite{Kulicrev}. It was found that at weak coupling
the Lorentzian FSP led to increase of $T_{c}$, while at strong coupling the
phonon contribution was detrimental due to large renormalization parameter.
Consensus emerged that the EPI of $CuO$ plane phonons alone is not strong
enough to get such a high $T_{c}$. EPI \ exchange can somewhat enhance, but
cannot be the major cause of the d-wave pairing.

In view of the experience with 1UC $FeSe/STO$, is is natural to ask whether
the lateral apical oxygen phonon exchange that \textit{naturally has}
exponential FSP, due to distance $d_{a}$ between the conducting and
insulating layers, can lead to the \textit{d-wave} pairing in cuprates. It
immediately reminds a high $T_{c}$ "smoking gun" that was observed of more
than a decade ago. It was discovered\cite{smokinggun} that the
superconducting gap in $Bi2212$ is (locally) anti- correlated precisely to
the distance, $d_{a}$, between the $Cu$ atoms and the apical oxygen atoms
just below/above. This is the first "smoking gun" pointing at crucial role
of the apical oxygen atoms. The evidence of the anti - correlation is not
conclusive since recently correlation single-layer $%
Bi_{2}Sr_{0.9}La_{1.1}CuO_{6}$, double-layer $Nd_{1.2}Ba_{1.8}Cu_{3}O_{6}$
and infinite-layer $CaCuO_{2}$ was observed \cite{Peng17}.

The second smoking gun is the tunneling experiment\cite{DavisBalatsky} that
the authors describe best: "We find intense disorder of electron - boson
interaction energies at the nanometer scale, along with the expected
modulations in $d^{2}I/dV^{2}$. Changing the density of holes has minimal
effects on both the average mode energies and the modulations, indicating
that the bosonic modes are unrelated to electronic or magnetic structure.
Instead, the modes appear to be local lattice vibrations, as substitution of 
$^{18}O$ for $^{16}O$ throughout the material reduces the average mode
energy by approximately $6\%$ - the expected effect of this isotope
substitution on lattice vibration frequencies." This is an indication that
vibrating oxygen atoms are out of the $CuO$ plane. We therefore revisit this
clear evidence in light of the lateral apical vibration superconductivity
theory.

Unlike 1UC $CuO$, where no measurements of the phonon excitations were made
to date, the bulk $BSCCO$ crystals were thoroughly studied. Evidence
consists of the "kink" in quasiparticle dispersion relation in normal state%
\cite{kink1,Lanzara04,kink2} measured by ARPES, large isotope effect
observed mainly in underdoped samples\cite{BSCCOisotope} and the statistics
of the STM measurements\cite{DavisBalatsky}. The kinks should be attributed
to EPI, since their locations (energies) change\cite{kink2} by 6\% upon
substitution of the $^{16}O$ isotope by $^{18}O$. The distribution of $%
d^{2}I/dV^{2}$ is independent of doping in a wide range. In particular its
average value is $40mev$ and is shifted by 6\% upon the isotope substitution%
\cite{DavisBalatsky}. This indicates that if the phonon pairing mechanism is
dominant the relevant phonons do not belong to the $CuO$ planes. Phonons in
cuprates were extensively studied within the microscopic (DFT) approach
including the oxygen vibration mode\cite{Falter}.

In the present paper we construct a theory of a high $T_{c}$ cuprate that
based on the idea of dominant pairing due to \textit{apical lateral
longitudinal } \textit{phonons }(ALLP) along with minor AF fluctuations
contribution. This $40mev$ phonon mode and its coupling including the matrix
elements are described sufficiently well by the Born - Meyer approximation%
\cite{Abrahamson},\cite{averestov} that has been applied to cuprates\cite%
{Falter93}. To support the pairing scenario, it is crucial to present a
simple enough microscopic model of cuprates that comprehensively describes
(at least qualitatively) various features of \ the material over the whole
doping - temperature phase diagram (underdoped to overdoped) including both
normal and d -wave superconducting states. To be more specific we consider
the effect of the ALLP pairing in the arguably best studied cuprate
superconductor $Bi2212$. To describe the pseudogap physics of 2DEG in the $%
CuO$ planes we limit ourselves to the fourfold symmetric $t-t^{\prime }$
single band Hubbard model\cite{Dagotto1} with on site repulsion energy $U$.
In the absence of direct experimental determinations of $U$, one resorts to
the first principle calculations. Most of the microscopic (DFT)
determinations of $U$ \cite{DFTlargeU} are in the "strong coupling Mott
insulator" range $U/t=5-10$, so that $U$ is comparable to the bandwidth $W$.
Recently however in a similar type of the first principle calculations\cite%
{DFTsmallU} resulted in smaller values of $U$. It turns out within our
approach that in order to describe the pseudogap physics, parameters of the
model are restricted to a rather narrow "window" around $t^{\prime }\sim
-0.2t$, $U\sim 6t$. Since the ALLP exchange is effective enough to be the
dominant "glue" responsible for the d - wave pairing, a simple description
of the Hubbard model combining the RPA type coupling renormalization due to
screening\cite{Maier20} and the symmetrized HF approach\cite{Li19} is
sufficiently accurate. The spin fluctuations exchange enhances
superconductivity by 15-20\%.

Two conditions turned out to be sufficient to trigger robust apical phonon d
- wave pairing: the rhombic shape of the Fermi surface and the exponential
FSP of the ALLP mode. The dependence of the superconducting gap on doping,
temperature and effect of the $^{16}O\rightarrow ^{18}O$\ isotope
substitution are obtained. In normal state the dimensionless EPI strength is 
$\lambda \sim 0.6$, thus justifying the use of the weak coupling approach%
\cite{McMillan68}. The phonons naturally explain the effect of the isotope
substitution on the kink in dispersion relation.

The paper is organized as follows. In Section II a sufficiently precise
phenomenological model of the lateral optical phonons in ionic crystal is
developed. In Section III an effective model of the correlated electron gas
is presented. Section IV is devoted to normal state properties: the
pseudogap phenomena (including the $T^{\ast }\,$line, fragmentation of the
quasi - particle spectrum) and renormalization of the electron Green's
function due to phonons. This allows location of kink in dispersion relation
(including the isotope dependence) and the EPI coupling $\lambda $. In
Section V superconductivity is studied in the framework of dynamic
Eliashberg approach. Both the phonon and the spin fluctuation channels are
accounted for over the full doping range. The isotope effect exponent is
determined. In the last Section results are summarized and discussed. A
simplified general picture of the d - wave pairing by apical phonons and its
coexistence with spin fluctuations is presented.

\section{The model}

Our model consists of the 2DEG interacting with phonons of a polar insulator:%
\begin{equation}
H=H_{Hub}+H_{ph}+H_{e-ph}\text{.}  \label{Hamiltoniandef}
\end{equation}%
We start with the phonon. The electron part is the Hubbard model, while the
coupling between the electronic and vibrational degrees of freedom, $%
H_{e-ph} $, is subject of the last Subsection.

\subsection{What phonons are contributing most to the electron - electron
pairings?}

Although the prevailing hypothesis is that superconductivity in cuprate is
"unconventional", namely not to be phonon - mediated, the phonon based
mechanism has always been a natural option to explain extraordinary
superconductivity in cuprates. As mentioned in Introduction, the most
studied phonon glue mode has been the oxygen vibrations within the $CuO$
plane\cite{Bulut}\cite{Annett}\cite{Kulicrev}\cite{Kulichearly}. As argued
in ref.\cite{Rosen19}, in the context of high $T_{c}$ 1UC $FeSe$ on
perovskite substrates, lateral vibrations of the oxygen atoms in the
adjacent ionic perovskite layer can couple sufficiently strongly to 2DEG
residing in the $CuO$ plane to be a viable option. Qualitatively one of the
reasons is that the $SrO$ layer constitutes a strongly coupled ionic
insulator. Unlike the metallic layer where screening is strong, in an ionic
layer screening is practically absent and a simple microscopic theory of
phonons and their coupling exists\cite{Abrahamson}. It was repeatedly noticed%
\cite{Gorkov} that vibrations in $c$ directions contribute little to
pairing. Let us start with a brief description of the structure of the
perhaps best studied high $T_{c}$ material $Bi2212$. Then the microscopic
lateral vibrations model is presented, while their coupling to the electron
gas is considered in the next Section.

\begin{figure}[h]
\centering \includegraphics[width=8cm]{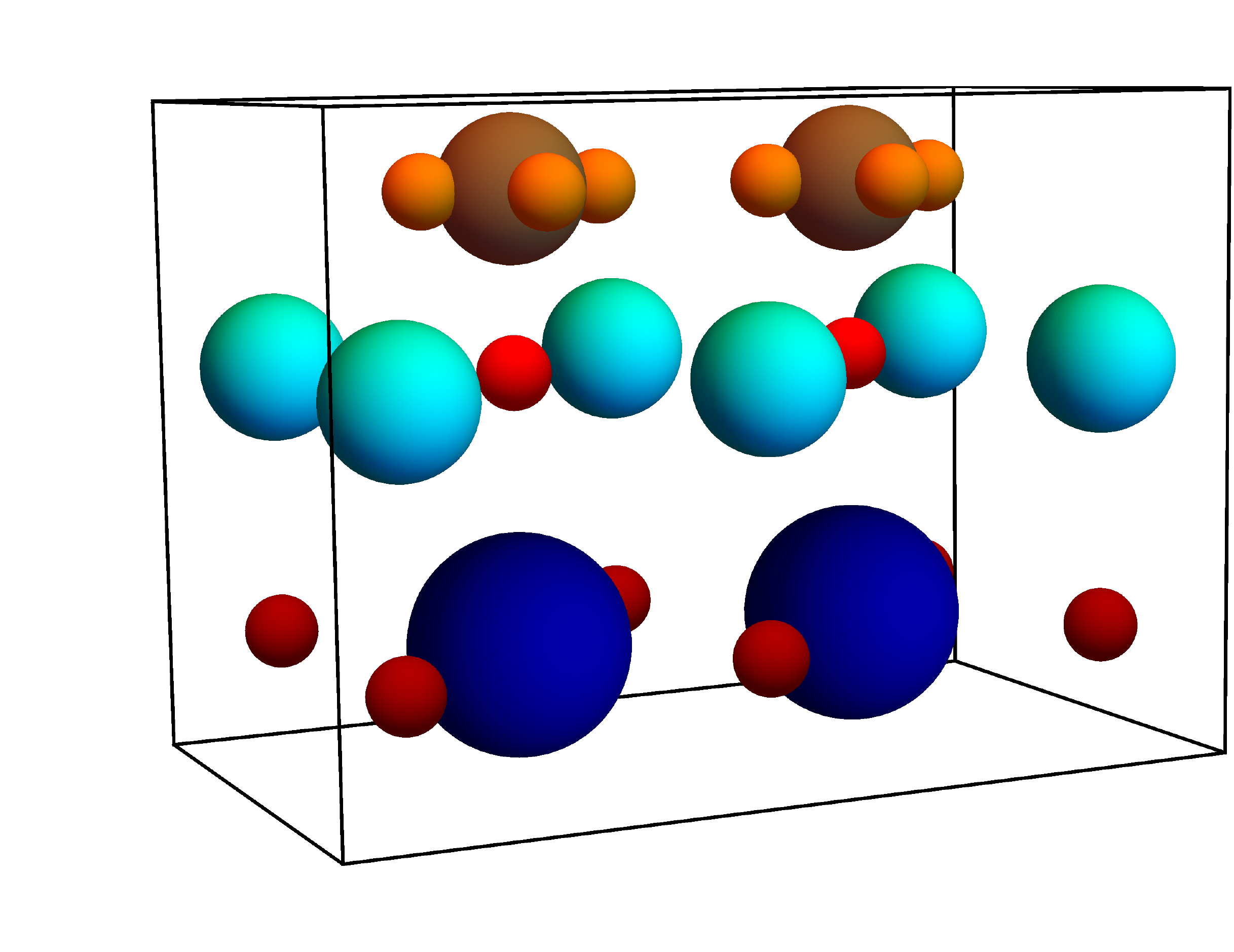}
\caption{The profile 3D view of three layers comprising relevant part of the
one unit cell :molecule" of $Bi2212$. Top (2DEG) layer: $Cu$ (brown) $O_{2}$
(orange), the apic phonon layer: $Sr$ (cyan) $O$ (red). The third layer: $Bi$
(violet) $O$ (dark red). Sizes of atoms are inversely proportional to the
values of the Born - Mayer inter - atomic potential parameter parameter $b$
in Eq.(\protect \ref{interatomic}).}
\end{figure}

\begin{table*}[h]
\caption{Atomic parameters determining lateral apical oxygen vibrations.}
\begin{center}
\begin{tabular}{|c|c|c|c|c|c|c|}
\hline
$\text{atom}$ & $Cu$ & $O_{1}$ & $Sr$ & $O_{2}$ & $Bi$ & $O_{3}$ \\ \hline
$\text{mass (a.u.)}$ & $64$ & $16$ & $88$ & $16$ & $209$ & $16$ \\ \hline
$A\, \  \text{(}kev\text{)}$ & $13.919$ & $2.143$ & $20.785$ & $2.143$ & $%
63.922$ & $2.143$ \\ \hline
$b$ \ ($A^{-1}$) & $3.561$ & $3.788$ & $3.541$ & $3.788$ & $3.4998$ & $3.788$
\\ \hline
charge $Z$ & $2.4$ & $-1.2$ & $.95$ & $-.95$ & $1.33$ & $-1.33$ \\ \hline
spacing $z$ ($A\,$) & $1.84$ & $1.84$ & $0$ & $0$ & $-2.75$ & $-2.75$ \\ 
\hline
\end{tabular}%
\end{center}
\end{table*}

The structure of the quarter of the $Bi_{2}Sr_{2}CaCu_{2}O_{8+\delta }$ unit
cell near the conducting layer is schematically depicted in Fig. 1.
Electronic properties in both normal and superconducting states of cuprates
are determined by holes (created by doping) in conducting $CuO$ layers, see
top layer in Fig.1 (where $Cu$ is drawn as a brown sphere, $O$ - small
orange spheres) and the left most chart in Figs. 8 (Appendix A). Besides the
single $CuO_{2}$ layer only two insulating oxide layers are assumed to be
relevant. The closest layer at distance $d_{a}=1.84A$, see the second chart
from left in Fig.8a, consists of heavy $Sr$ atoms (cyan rings) and light
"apical" oxygen (small red circle). The next layer is $BiO$, see the third
chart from left in figure in Fig. 8b ($Bi$ - violet large ring, $O$ - small
dark red circles). Below this layer the pattern is replicated in reverse
order. Of course $Bi2212$ has metallic bilayers separated by $Ca$. In this
paper we neglect the effects of tunneling between the $CuO_{2}$ layers. Out
of plane spacings counted from the $SrO$ layer are specified in Table I.

The translational symmetry in the lateral ($x$,$y$) directions of the system
has the lattice spacing of $a=3.9A$ and coincides with the distance between
the $Cu$ atoms. Distances between the layers are also given Table I
neglecting small canting. The crystal has very rich spectrum of phonon
modes. However very few have a strong coupling to 2DEG and even fewer can
generate lateral (in plane) forces causing pairing. While phonons within the 
$CuO$ planes have been extensively studied both theoretically\cite%
{Bulut,Kulicrev} and experimentally, the conclusion is that they do not
constitute a strong enough "glue". It is reasonable to expect that the modes
most relevant for the electron - phonon coupling are the vibrations of the
atoms in the adjacent $SrO$ layer, see Fig.1. This is in conformity with the
first and second "smoking gun" experiment findings\cite{smokinggun}\cite%
{DavisBalatsky}: the "glue" is independent of the doping and anything else
that happens in the 2DEG in the $CuO_{2}$ layer simply because the phonons
are originating in different layer.

\subsection{Lateral apical oxygen optical phonon modes in the $SrO$\ layer.}

Phonons in ionic crystals are described by the Born - Meyer potential due to
electron's shells repulsion\cite{Abrahamson} and electrostatic interaction
of ionic charge,%
\begin{equation}
V^{XY}\left( r\right) =\sqrt{A_{X}A_{Y}}\exp \left[ \frac{1}{2}\left(
b^{X}+b^{Y}\right) r\right] +Z_{X}Z_{Y}\frac{e^{2}}{r}\text{,}
\label{interatomic}
\end{equation}%
with values of coefficients $A$ and $b$ listed in Table I. The ionic charges 
$Z$ are estimated from the DFT calculated Milliken charges\cite{averestov}.
In the $SrO$ layer the charges are constrained by neutrality. Since oxygen
is much lighter than $Sr$, the heavy atoms' vibrations are negligible.
Obviously that way we lose the acoustic branch, however it is known that the
acoustic phonons contribute little to the pairing\cite{Mahan,Gorkov}. Atoms
in neighboring layers can also be treated as static. Moreover one can
neglect more distant layers. Even the influence of the lower $BiO$ layer
(below the last layer shown in Fig.1) is insignificant due to the distance.
Consequently the dominant lateral displacements, $u_{\mathbf{m}}^{\alpha }$, 
$\alpha =x,y$, are of the oxygen atoms directly beneath the $Cu$ sites at $%
\mathbf{r}_{\mathbf{m}}=a\left( m_{1},m_{2}\right) $.

The dynamic matrix $D_{\mathbf{q}}^{\alpha \beta }$ is calculated by
expansion of the energy to second order in oxygen displacement (details in
Appendix A), so that the phonon Hamiltonian in harmonic approximation is:

\begin{equation}
H_{ph}=\frac{1}{2}\sum \nolimits_{\mathbf{q}}\left \{ M\frac{du_{-\mathbf{q}%
}^{\alpha }}{dt}\frac{du_{\mathbf{q}}^{\alpha }}{dt}+u_{\mathbf{-q}}^{\alpha
}D_{\mathbf{q}}^{\alpha \beta }u_{\mathbf{q}}^{\beta }\right \} \text{.}
\label{Hph}
\end{equation}%
Here $M$ is the oxygen mass. Summations over repeated components indices is
implied. Now we turn to derivation of the phonon spectrum. Two eigenvalues,
the transversal (red) optical (TO) and the longitudinal (blue) optical (LO)
modes are given in Fig. 2. One observes that there are longitudinal modes
are in the range $\Omega _{\mathbf{q}}\sim $ $26-41mev$ and $22-32mev$
respectively. The energy of LO modes is larger than that of the
corresponding TO, although the sum $\Omega _{\mathbf{q}}^{LO}+\Omega _{%
\mathbf{q}}^{TO}$ is nearly dispersionless. At $\Gamma $ the splitting is
small, while due to the long range Coulomb interaction there is hardening of
LO and softening of TO at the BZ edges. The dispersion of the high frequency
modes is small, while for the lower frequency mode it is more pronounced.

\begin{figure}[h]
\centering \includegraphics[width=14cm]{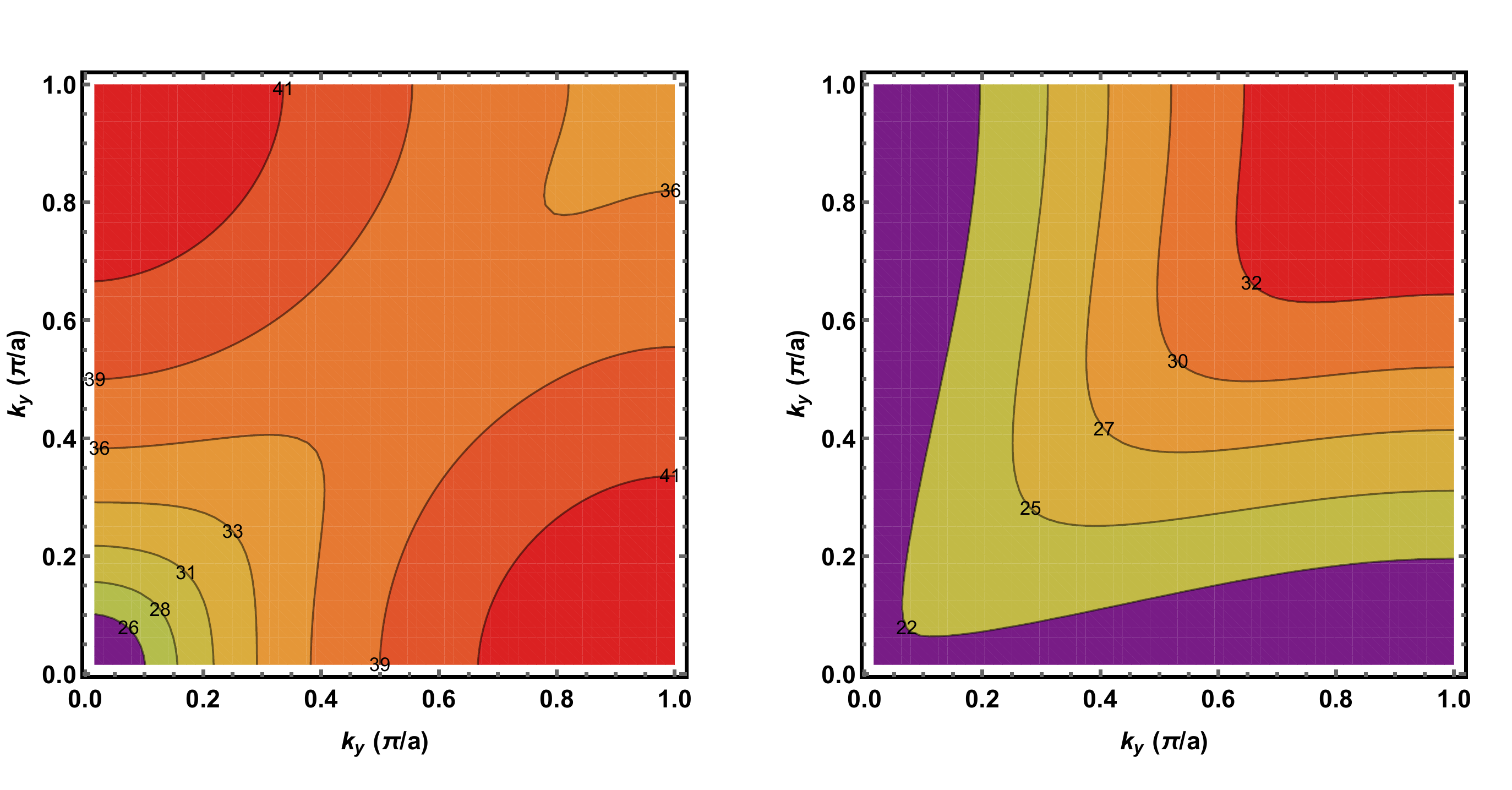}
\caption{ Spectrum of the lateral apical oxygen vibrations in the $SrO$
plane. a. longitudinal optical modes, b. transverse optical modes. Note
moderate dispersion of the longitudinal mode.}
\end{figure}

\subsection{The $t-t^{\prime }$ Hubbard model of the 2DEG in $CuO$ layers.}

The electron gas of $Bi2212$ consists of two identical layers with tunneling
between them. The effective single band model. Neglecting the inter - layer
tunneling, the simplest $t-t^{\prime }$ Hamiltonian in momentum space is:

\begin{equation}
K=\sum \nolimits_{\mathbf{k}}c_{\mathbf{k}}^{\sigma \dagger }\left( \epsilon
_{\mathbf{k}}+\epsilon _{\mathbf{k}}^{\prime }-\mu \right) c_{\mathbf{k}%
}^{\sigma }\text{,}  \label{tight1}
\end{equation}%
Here $c_{\mathbf{k}}^{\sigma \dagger }$ is the electron creation operator
with spin projection $\sigma =\uparrow ,\downarrow $ . Only nearest and next
to nearest neighbors hopping terms are included: 
\begin{eqnarray}
\epsilon _{\mathbf{k}} &=&-2t\left( \cos \left[ ak_{x}\right] +\cos \left[
ak_{y}\right] \right) ;  \label{epsilondef} \\
\text{ }\epsilon _{\mathbf{k}}^{\prime } &=&-4t^{\prime }\cos \left[ ak_{x}%
\right] \cos \left[ ak_{y}\right] \text{.}  \notag
\end{eqnarray}%
Summations are always over the 2D Brillouin zone, $-\pi /a<k_{x},k_{y}<\pi
/a $. The dispersion relation thus is simplified with respect to a
"realistic" one\cite{Kordyuk},\cite{Tstar} in which splitting due to
tunneling is also taken into account and more distant hops are included.
Values of the hopping parameters, see Table II will be fixed independently
of chemical potential $\mu $ determining the (hole) doping $x$. Reasons for
such a choice will be given after the phase diagram will be presented in the
next Section.

The on site repulsion is described by the on site Hubbard repulsion term\cite%
{Dagotto1}%
\begin{equation}
V=U\sum \nolimits_{\mathbf{i}}n_{\mathbf{i}}^{\uparrow }n_{\mathbf{i}%
}^{\downarrow }\text{,}  \label{Vdef}
\end{equation}%
with $n_{\mathbf{i}}^{\sigma }=c_{\mathbf{i}}^{\sigma \dagger }c_{\mathbf{i}%
}^{\sigma }$ being the spin $\sigma $ occupation on the site $\left \{
i_{x},i_{y}\right \} $. Due to strong repulsion, even the model without
phonons is highly nontrivial and will be treated approximately in the next
Section. Now we turn to the electron - phonon coupling.

While the lattice spacing $a$ is firmly determined by experiment (and is
nearly independent of doping for small $x$), the microscopic \cite{DHLee} or
phenomenological\cite{Kordyuk} estimates for other electron gas parameters
like the energy scales $U,t,t^{\prime },\mu $ vary considerably in different
one band Hubbard approaches. The values of $t=0.3eV$, $U=6$ at zero doping
will be used throughout the paper to fit numerous experimental quantities
like the ARPES\cite{Kordyuk}, the pseudogap characteristics\cite{Tstar}. The
range of acceptable values of $t^{\prime }/t$ is rather limited. If one
chooses $\left \vert t^{\prime }\right \vert /t<0.12$, the Mott state at
very low doping does not appear\cite{Irkhin16}. At values larger than $%
\left
\vert t^{\prime }\right \vert /t>0.25$ the shape of the Fermi surface
in the underdoped regime is qualitatively different from the one observed by
ARPES\cite{ARPES-Ding18}. The value of $t^{\prime }=-0.17t$ is chosen to
tune the Lifshitz (topological) transition from the full Fermi surface to
the fractured one (four arcs) occurs at experimentally observed\cite%
{accurate} doping $x^{opt}=0.16$.

\subsection{Electron - phonon coupling}

The lateral apical oxygen phonon's interaction with the 2DEG on the adjacent 
$CuO$ layer $d_{a}=1.84A$ above the $SrO$ plane is determined by the
electric potential created the charged apical oxygen vibration mode $\mathbf{%
u}_{\mathbf{m}}$ at arbitrary point $\mathbf{r}$ is:

\begin{equation}
\Phi \left( \mathbf{r}\right) =\sum \nolimits_{\mathbf{m}}\frac{Ze}{\sqrt{%
\left( \mathbf{r}-\mathbf{r}_{\mathbf{m}}-\mathbf{u}_{\mathbf{m}}\right)
^{2}+d_{a}^{2}}}\text{,}  \label{3_pot}
\end{equation}%
Here the apical oxygen charge taken to be $Z=-0.95$, see Table I. This value
is slightly below the charge at which transition to charge density wave
occurs. The interaction electron-phonon Hamiltonian that accounts for the
hole charge distribution in the $CuO$ plane is derived in Appendix A. The
result in momentum space has a density - displacement form

\begin{equation}
H_{eph}=Ze^{2}\sum \nolimits_{\mathbf{q}}n_{-\mathbf{q}}g_{\mathbf{q}%
}^{\alpha }u_{\mathbf{q}}^{A\alpha }\text{,}  \label{Heph}
\end{equation}%
with EPI matrix element, 
\begin{equation}
\mathbf{g}_{\mathbf{q}}=\frac{1}{2}\left( \cos \frac{aq_{x}}{2}+\cos \frac{%
aq_{y}}{2}\right) \overline{\mathbf{g}}_{\mathbf{q}};\text{ \  \ }\overline{%
\mathbf{g}}_{\mathbf{q}}\approx 2\pi e^{-qd_{a}}\frac{\mathbf{q}}{q}\text{.}
\label{A}
\end{equation}%
It is well known that only longitudinal phonons contribute to the effective
electron - electron interaction, as is clear from the scalar product form of
the Eq.(\ref{Heph}). The precision of the last equality is 2\%, see figure
9,10 in Appendix A.

To conclude Eqs.(\ref{tight1},\ref{Hph},\ref{Heph}) define our microscopic
model. Now we turn to description of the normal state properties of 2DEG,
including the influence of the EPI.

\section{Normal state properties: pseudogap, EPI coupling strength and kink
in dispersion relation.}

The normal state of cuprates exhibits a host of phenomena including
pseudogap in underdoped regime resulting in fracture of the Fermi surface,
significant charge and spin susceptibility due to strong anti -
ferromagnetic correlations (leading to enhancement of the d - wave pairing).
These phenomena are described in the framework of the strongly coupled
Hubbard model defined in the previous Section. Unfortunately the theoretical
description of the Hubbard model away from half filling (Monte Carlo \cite%
{Sorella}, diagrammatic\cite{Katzenelson,Held}) is either uncertain or
extremely complicated. We use a much simpler approximation scheme including
the RPA type coupling $U$ renormalization \cite{Maier07} and symmetrized HF%
\cite{Li19}. It provides a good agreement with the more sophisticated
methods. Coupling to phonons also affects the normal properties such as the
dispersion relation. The strength of EPI will be estimated and the quasi -
particle self energy calculated perturbatively.

\subsection{ Renormalized mean field description of the Hubbard model}

Hubbard model at moderate value $U=6$ in the doping range $x=0.05-0.25$
range is a strongly correlated fermion system that does not allow the Landau
liquid description (except at high doping). Generally it is also out of
applicability range of the HF approximation due to large vertex corrections 
\cite{Katzenelson,Held}. However it is well known that the overdoped system
has a well defined Fermi surface and can be very well described by the HF
type two - body correlator \cite{Kordyuk}. In the underdoped phase one
obtains an effective description in terms of "RVB" correlators\cite{ZhangRVB}
that have recently been cast as a symmetrized HF\cite{Li19}. Such an
approach is consistent if the vertex corrections effectively lead to
reduction of the coupling to a smaller value $\overline{U}$. It turns out
that MC and diagrammatic results can be approximated by such a scheme when
the renormalized $\overline{U}$,

\begin{equation}
\overline{U}=\frac{U}{1-\frac{U}{2}\chi _{0}},  \label{Ubar}
\end{equation}%
where $\chi _{0}$ is the (Matsubara) charge susceptibility. This should be
solved consistently with the HF equations and is described in both the
overdoped and the underdoped phases in Appendix B.

The coupling $U=6$ is reduced by screening to the renormalized values given
in Table II. The HF equations were solved numerically by iterations on
lattice $N=128\times 128$ with periodic boundary conditions.

\begin{table*}[h]
\caption{ Effective (renormalized) coupling as function of doping for bare
coupling $U=6\ $and $t^{\prime }/t=-0.17.$}
\begin{center}
\begin{tabular}{llllllllllllllllll}
hole doping $x(\%)$ & $1$ & $2$ & $3$ & $5$ & $7$ & $9$ & $11$ & $13$ & $14$
& $15$ & $16$ & $17$ & $19$ & $21$ & $23$ & $25$ & $28$ \\ 
eff. coupling $U_{r}$ & $4.35$ & $4.06$ & $3.93$ & $3.82$ & $3.67$ & $3.52$
& $3.33$ & $3.1$ & $2.94$ & $2.65$ & $1.91$ & $1.93$ & $2.03$ & $2.14$ & $%
2.23$ & $2.3$ & $2.33$%
\end{tabular}%
\end{center}
\end{table*}
One of the striking normal state phenomena in underdoped cuprates is
pseudogap\cite{pseudogapexp,pseudogap}. In the present paper we adopt a
point of view that pseudogap to the short range anti - ferromagnetic order
within each of the $CuO$ layers. The long range AF order is lost at a
relatively small doping and the system becomes quasi two dimensional. In 2D
one can model the short range order and the fluctuations effects\cite%
{Dagotto1} by considering the macroscopic sample as a system of AF domains
with certain domain size. Generally local (STM) probes described in
Introduction provide distribution of quantities like pseudogap within the
domains. On the other hand ARPES, thermodynamic and transport experiments
provide information on all the scales, namely after averaging over the
domains. It is found that the value of the pseudogap is qualitatively agree
with somewhat similar calculations\cite{Metzner07} (improved by the
renormalization group), the MC simulations\ and experiments \cite%
{pseudogapexp,pseudogap}.

The transition temperature $T^{\ast }$ as function of the hole doping, $%
x=1-n $, is given in Fig. 3 as the green line. It starts at the quantum
critical point $x^{\ast }=0.16,$ rapidly increases (almost vertically
although a slight bending is visible) intersecting with the superconducting
transition temperature $T_{c}$. Then it curves towards the AF phase at small
doping. The mean field transition happens to be second order with an
exception of the small section below the "superconducting dome" in Fig. 3
(marked by a phenomenological parabolic fit to experiment, see ref.\cite%
{accurate}).

\begin{figure}[h]
\centering \includegraphics[width=12cm]{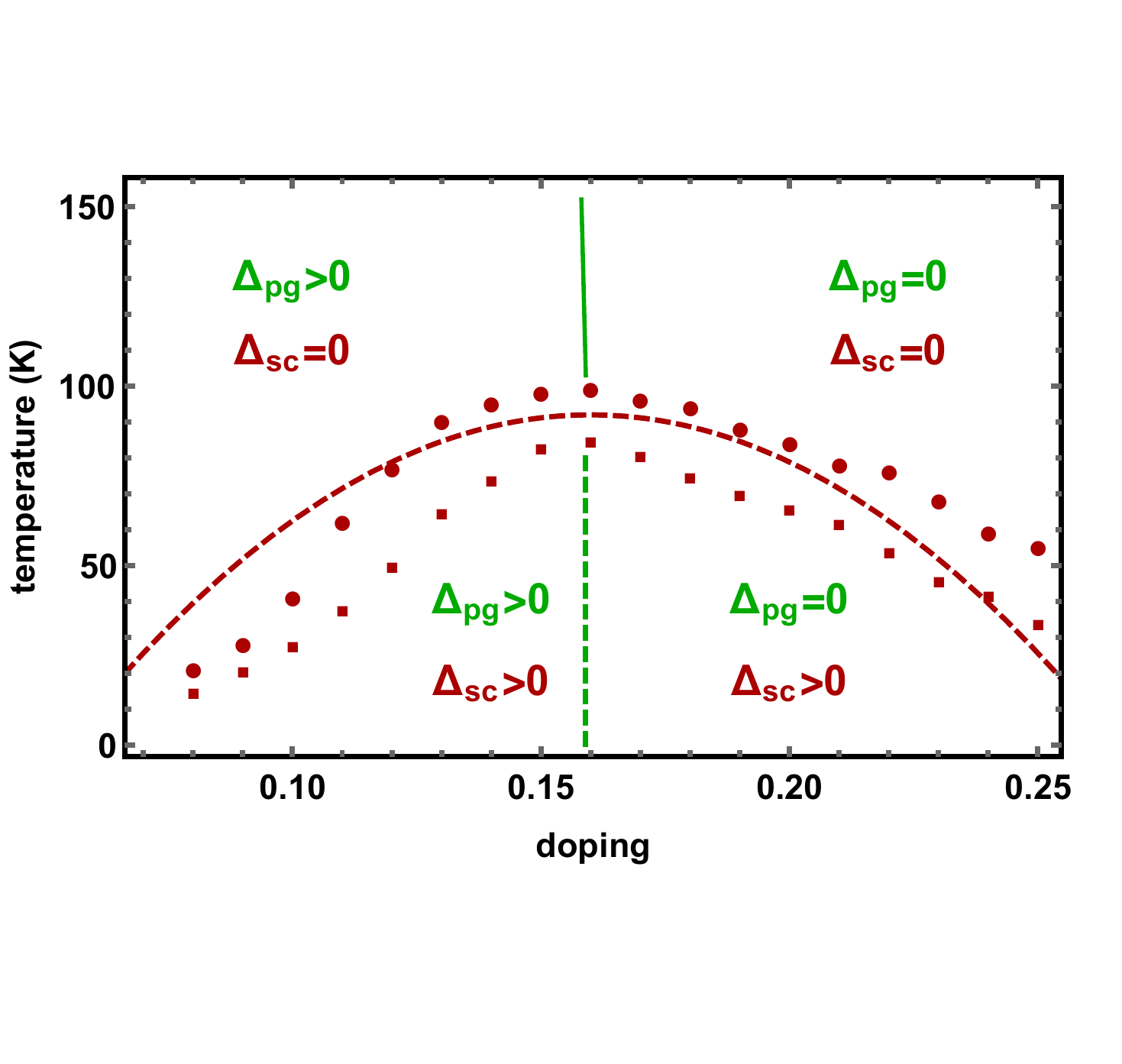}
\caption{The doping - temperature phase diagram of a hole doped cuprate. The
green curve marks the pseudogap transition $T^{\ast }$. Solid line
represents the (mean field) second order transition, while the dashed
segment represent weakly first order one, The parabolic curve is the
experimental superconductor - normal critical temperature in $Bi2212$
measured in ref. \protect \cite{accurate}. Red points are $T_{c}$ of our
model, while the blue points are critical temperatures due to the apical
phonon's pairing only (that is when the spin fluctuations are ignored).}
\end{figure}

In Appendix B the expressions for the electron correlators in both phases is
given. The spectral weight namely the imaginary part of the symmetrized
Green function, Eq.(\ref{AFcorr}), at zero frequency, exhibits the fractured
Fermi surface qualitatively similar to ARPES observation\cite%
{ARPES-Ding18,pocket}. 
\begin{figure}[h]
\centering \includegraphics[width=18cm]{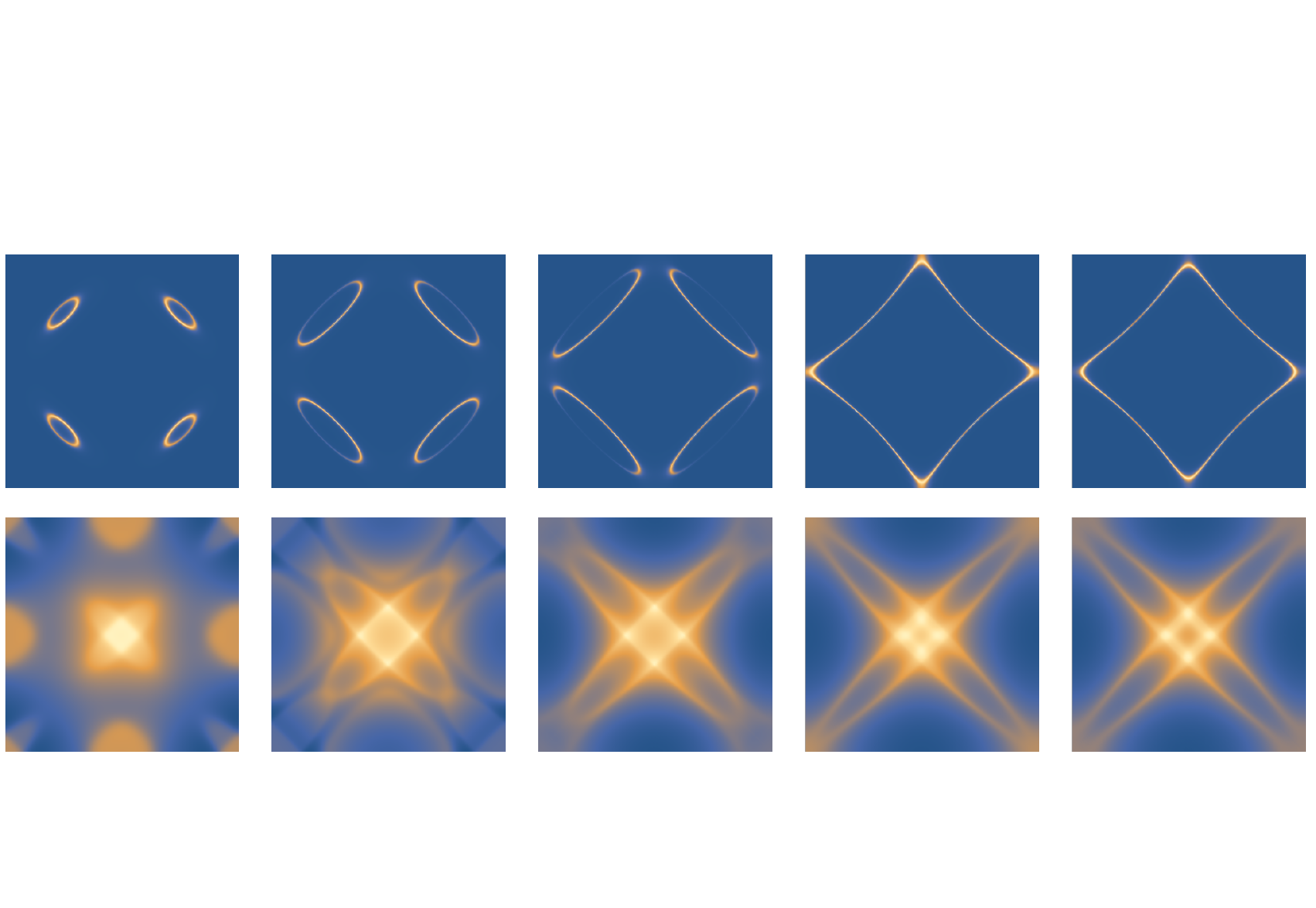}
\caption{Upper row: the quasi - particle spectral weight in
nonsuperconducting ($x=0.03$) underdoped ($x=0.11,0.15$), optimal doping ($%
x=0.16$)$,$ and overdoped ($x=0.18$) systems. Four Fermi arcs in underdoped
case coalesce into a closed Fermi surface at the (Lifshitz) topological
transition at optimal doping. Lower row: Spin susceptibility distribution of
\ (in $meV^{-1}$)\ for the same doping leveks. The distribution is
continuous through the Lifshitz transition at optical doping. Note that
Brillouin zone in the upper row is centered at the chrystallographic $\Gamma 
$ point, while in the lower row it is shifted to the $M$ point. This allows
a convenent focus on the peak around the AF order $\mathbf{Q}=\left( \protect%
\pi /a,\protect \pi /a\right) $ point.}
\end{figure}

The spectral weigh for five values of doping are shown in the upper row in
Fig.4. Two are in the non -superconducting state, $\ x=0.02,0.05$, one in
the underdoped region, $\ x=0.13$, optimal, $x=x^{opt}=0.16$, \ and
overdoped $x=0.2$ regions. One observes that as the doping increases the
length of the four Fermi arcs increases until the topological (Lifshitz)
transition to a single Fermi surface at $x^{opt}$. Upon further hole doping
the area of the enclosed region of BZ decreases. Note that the Fermi surface
does not extend to the BZ boundary as seen in early experiments\cite{Kordyuk}%
, however more recent measurements\cite{ARPES-Ding18} apparently are
consistent with this picture.

\subsection{Phonon renormalization of the quasi - particle self energy and
coupling constant $\protect \lambda ^{ph}$}

\subsubsection{Self energy due to phonons.}

The quasiparticle (HF) self - energy is renormalized due to interaction with
phonons. It generally leads to characteristic features of the spectrum like
satellite bands\cite{Lee12}\cite{Leerev}, kinks in dispersion relation\cite%
{kink1}\cite{kink2}, etc. at energies of the order of the phonon frequency $%
\Omega $ above and below Fermi level. In our case (for details see a more
general case considered in ref.\cite{Rosen19} and references therein) the
Matsubara self energy for $x>x^{opt}$ (and temperature above $T_{c}$) in
(gaussian or renormalized) perturbation theory is:%
\begin{equation}
\Sigma _{n\mathbf{k}}=\frac{\left( 2\pi Ze^{2}\right) ^{2}T}{MN^{2}}\sum
\nolimits_{\mathbf{l,}m}\frac{e^{-2ld_{a}}}{\omega _{m}^{b2}+\Omega ^{2}}%
\frac{1}{i\omega _{n+m}-E_{\mathbf{k+l}}}\text{,}  \label{SE}
\end{equation}%
where 
\begin{equation}
E_{\mathbf{p}}\equiv \epsilon _{\mathbf{p}}+\epsilon _{\mathbf{p}}^{\prime
}-\mu +Un/2.  \label{EHFdef}
\end{equation}%
The dispersion relations are given in Eq.(\ref{epsilondef}) and $M$ is the
oxygen ion mass. Second order "gaussian" perturbation theory\cite{gausspert}
is justified at weak coupling, so that it should be confirmed in the
following subsection that the dimensionless effective electron - electron
coupling $\lambda _{ph}$ is indeed small. Summing\ over\ the bosonic\
Matsubara\ frequencies,\ $\omega _{m}^{b2}=2\pi Tm$,\ one\ obtains (after
analytic continuation to physical frequency),

\begin{eqnarray}
\Sigma \left( \omega ,\mathbf{k}\right) &=&\frac{\left( 2\pi Ze^{2}\right)
^{2}}{2M\Omega N^{2}}\sum \nolimits_{\mathbf{l}}e^{-2d_{a}l}I_{\omega ,%
\mathbf{k+l}};  \label{Iover} \\
I_{\omega \mathbf{p}} &=&\frac{f_{B}\left[ \Omega \right] -f_{F}\left[ -E_{%
\mathbf{p}}\right] +1}{\omega +i\eta +\Omega -E_{\mathbf{p}}}+\frac{f_{B}%
\left[ \Omega \right] +f_{F}\left[ -E_{\mathbf{p}}\right] }{\omega +i\eta
-\Omega -E_{\mathbf{p}}}\text{,}  \notag
\end{eqnarray}%
where $f_{B}\left[ \varepsilon \right] =\left( \exp \left[ \varepsilon /T%
\right] -1\right) ^{-1}$ is the Bose distribution.

In the underdoped case ($x<x^{\ast }$) we make use of the symmetrized
correlators of the previous Subsection. The symmetrization is justified for
description of the ARPES data, since it is a nonlocal probe, presumably over
areas larger than the AF domain size. The results are similar in form to the
underdoped case:%
\begin{equation}
I_{\omega ,\mathbf{k}}=Z_{\mathbf{k}}^{+}\left( \frac{f_{B}\left[ \Omega %
\right] -f_{F}\left[ -E_{\mathbf{k}}^{+}\right] +1}{\omega +i\eta +\Omega
-E_{\mathbf{k}}^{+}}+\frac{f_{B}\left[ \Omega \right] +f_{B}\left[ -E_{%
\mathbf{k}}^{+}\right] }{\omega +i\eta -\Omega -E_{\mathbf{k}}^{+}}\right)
+\left \{ Z^{+},E^{+}\rightarrow Z^{-},E^{-}\right \} \text{.}
\end{equation}%
Here energies $E_{\mathbf{k}}^{\pm }$ and weights $Z_{\mathbf{k}}^{\pm }$
are given in Eqs.(\ref{Gsym},\ref{Enonmag}). These expressions will be used
for calculation of both the electron phonon coupling constant and the
dispersion relation\ of quasi - particles.

\subsubsection{Dimensionless electron - electron coupling $\protect \lambda $}

Generally the dimensionless coupling constant is defined in terms of the
self energy as $\lambda _{\mathbf{k}}=-\frac{d}{d\omega }\Sigma \left(
\omega ,\mathbf{k}\right) |_{\omega =0^{+}}$. In the overdoped case (see
Appendix B for details and expressions in a more cumbersome underdoped case)
one obtains at zero temperature: 
\begin{equation}
\lambda _{\mathbf{k}}=\frac{2\left( \pi Ze^{2}\right) ^{2}}{M\Omega N^{2}}%
\sum \nolimits_{\mathbf{l}}e^{-2d_{a}l}\left \{ \frac{\theta \left[ -E_{%
\mathbf{k+l}}\right] }{\left( E_{\mathbf{k+l}}-\Omega \right) ^{2}}+\frac{%
\theta \left[ E_{\mathbf{k+l}}\right] }{\left( E_{\mathbf{k+l}}+\Omega
\right) ^{2}}\right \} \text{.}  \label{lambda}
\end{equation}

Results of numerical computation at the nodal point on the Fermi surface in
the doping range from $x=0.08$ to $x=0.28$ is performed. At each doping the
location of the Fermi surface point was given by an analytic solution. As
expected it has a \ maximum of $\lambda ^{ph}=0.62$. Upon deviation from the
angle $45^{\circ }$ the coupling decreases. This is consistent with the
experimental value estimated recently\cite{Shenlambda}\cite{ScalapinoNat07}
at $30K$ to be $\lambda ^{ph}=0.41$ at optimal doping at \thinspace $\mathbf{%
k}=\left( 0.\pi \right) $. In the underdoped cases it vanishes at small
angles due to finite extent of the Fermi arc, Generally the averaged over
the Fermi surface coupling constant belongs to an intermediate range\cite%
{McMillan68}. Such coupling is sufficient (as will be shown also in the next
Section) to provide high d-wave superconductivity $T_{c}\sim 80-90K$ at
optimal doping, yet does not require the use of a rather problematic strong
coupling Eliashberg theory. \ The coupling constitutes the bulk of the
mechanism of superconductivity in the present paper (in addition to phonons
the spin fluctuations also contribute to the overall effective coupling $%
\lambda $, see below).

The EPI renormalizes the quasiparticle spectrum and dynamics leading to
several observations of the isotope substitution effect on the normal state
properties. One of them is the "kink" in dispersion relation. 
\begin{figure}[h]
\centering \includegraphics[width=12cm]{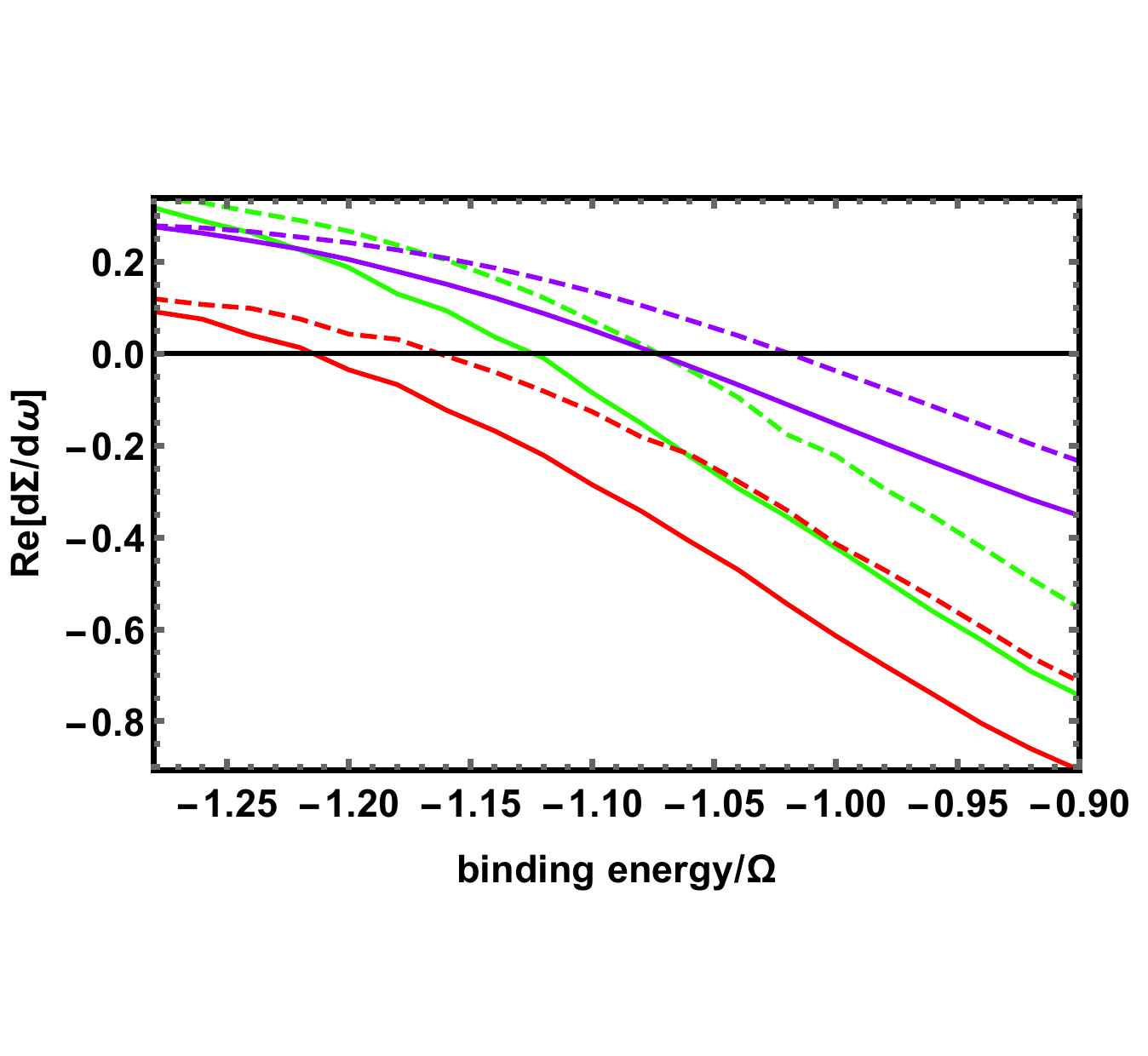}
\caption{Derivative of the self energy with respect to frequency at energies
around \ $\Omega $\ below the Fermi level. The values of doping are x=0.13
(green), x=0.15 (red) and x=0.17 (violet). The dash lines demonstrate
isotope effect when $^{16}O$ is replaced by $O^{18}$ isotope. The kink
appeares whe the derivative vanishes (blue line).}
\label{Fig5}
\end{figure}

\subsection{The "kink" function and\ the effect of the isotope substitution}

It was established by ARPES early on that the hole dispersion relation
abruptly changes derivative ("kink") in normal state approximately $45meV$
below Fermi level\cite{kink1,Lanzara04,kink2}. Although some other theories
appeared, the large isotope effect \cite{BSCCOisotope} (substitution of $%
^{16}O$isotope by $^{18}O$), observed mainly in underdoped samples) provides
evidence that he kinks should be attributed to EPI. To determine the kink
position observed directly, let us differentiate the self energy Eq.(\ref%
{Iover}) with respect to frequency $\omega $. The real part of the integrand
is:

\begin{equation}
\frac{d}{d\omega }I_{\omega ,\mathbf{p}}=-\frac{f_{B}\left[ \Omega \right]
-f_{F}\left[ -E_{\mathbf{p}}\right] +1}{\left( \omega +i\eta +\Omega -E_{%
\mathbf{p}}\right) ^{2}}-\frac{f_{B}\left[ \Omega \right] +f_{F}\left[ -E_{%
\mathbf{p}}\right] }{\left( \omega +i\eta -\Omega -E_{\mathbf{p}}\right) ^{2}%
}\text{,}  \label{B1}
\end{equation}%
where $E_{\mathbf{p}}$ was defined in Eq.(\ref{EHFdef}). In the underdoped
regime the expression is given in Appendix B.

To characterize the kink in dispersion relation, we calculate the derivative
in range of frequencies between $-1.3\Omega $ to $-0.9\Omega $ for three
dopings, $0.13$ (green), $0.15$ (red) and $0.17$ (violate) in Fig.5. The
kink position (zero value of the derivative) is around $\omega =-\Omega
=-45meV$ . The dashed lined are the same quantity but for a heavier isotope $%
^{18}O$, namely with the oxygen atom mass $M$ replaced by $\alpha M$, $\
\alpha =18/16$. The location is shifted by approximately $6\%$, as was
indicated in the ARPES experiment\cite{Lanzara04}. Now we turn to the main
objective of the present study: d - wave superconductivity.

\section{Superconductivity.}

Although the main emphasis of the paper is on the ALLP mechanism of the d -
wave superconductivity in the hole doped cuprates, in the present Section we
take into account also the magnetic fluctuation contribution. The reason is
that the AF fluctuations were widely observed and in certain cases were
shown to at least enhance superconductivity. The purpose of the present
Section is to quantitatively compare the role of these two contributions and
show how they coexist (complement each other) in the d - wave
superconducting state. We start from the derivation of the phonon exchange d
wave "potential" (mainly near the $\Gamma $ point of BZ) and then proceed to
the spin fluctuation one (mainly near the $M$ point of the BZ).

\subsection{Effective phonon and the spin fluctuation generated electron -
electron interactions in spin singlet channel}

In order to describe superconductivity, one should "integrate out" the
phonon and the spin fluctuations degrees of freedom to calculate the
effective electron - electron interaction. We start with the phonons. The
Matsubara action for EPI, Eq.(\ref{Heph}), and phonons, Eq.(\ref{Hph}), are,%
\begin{equation}
\frac{1}{T}\sum \nolimits_{m\mathbf{,q}}\left( Ze^{2}n_{-m,-\mathbf{q}}g_{%
\mathbf{q}}^{\alpha }u_{m,\mathbf{q}}^{\alpha }+\frac{M}{2}u_{-m,-\mathbf{q}%
}^{\alpha }\Pi _{m,\mathbf{q}}^{\alpha \beta }u_{m\mathbf{,q}}^{\beta
}\right) \text{,}  \label{action}
\end{equation}%
where $n_{-n,-\mathbf{q}}=\sum \nolimits_{\mathbf{k,}m}\psi _{\mathbf{k}-%
\mathbf{q},m-n}^{\ast \sigma }\psi _{\mathbf{k},m}^{\sigma }$ and $g$ was
defined in Eq.(\ref{A}). The polarization matrix is defined via the dynamic
matrix of Eq.(\ref{Hph}): $\Pi _{n,\mathbf{q}}^{\alpha \beta }=\left( \omega
_{n}^{b}\right) ^{2}\delta _{\alpha \beta }+M^{-1}D_{\mathbf{q}}^{\alpha
\beta }$, $\alpha ,\beta =x,y,$ calculated in Appendix A. Since the action
is quadratic in the phonon field $\mathbf{u}$, the partition function is
gaussian and can be integrated out exactly, see details in ref.\cite{Rosen19}%
. As a result one obtains the effective density - density interaction term
for of electrons

\begin{equation}
\mathcal{A}_{eff}^{ph}=\frac{1}{2T}\sum \nolimits_{\mathbf{q},n}n_{n,\mathbf{%
q}}v_{n\mathbf{q}}^{ph}n_{-n,-\mathbf{q}};\text{ \ }v_{n,\mathbf{q}}^{ph}=-%
\frac{\left( 2\pi Ze^{2}\right) ^{2}}{M}\frac{e^{-2d_{a}\left \vert \mathbf{q%
}\right \vert }}{\omega _{n}^{b2}+\Omega ^{2}}\text{.}  \label{Aee}
\end{equation}%
The expression is "exact" for harmonic phonons (we have neglected the
transversal mode and small dispersion of the longitudinal mode spectrum\cite%
{Rosen19}, see Fig. 2). An approximate expression for the effective
interaction due to the electron correlations effects will be derived next.
The potential exhibits the central "inverted" (that is negative) "peak" that
we will call the apical phonon dip due to the exponential form of the matrix
element. The second bosonic "glue" is generated by the correlation effects.

Since, as explained in Subsection IIIA, the renormalized on site repulsion
constant in our scheme, $\overline{U}$ is not very large (see Table III),
the gaussian expansion\cite{Li19}\cite{DMRTsmall} is applicable. One starts
with the mean field GF and considers the rest of the action as a
perturbation. In the overdoped case, it is just a "renormalized"
Kohn-Luttinger perturbation theory\cite{Kohn}. We therefore calculate the
effective interaction due to correlations in the second order in $U_{r}$.
Generally, utilizing the inversion symmetry, the effective interaction in
the spin singlet channel has a form: 
\begin{equation}
\mathcal{A}_{eff}^{cor}=\frac{1}{2T}\sum \nolimits_{\mathbf{q}.n}n_{n\mathbf{%
q}}v_{n\mathbf{q}}^{cor}n_{-n,-\mathbf{q}};\text{ \ }v_{m\mathbf{q}%
}^{cor}=U_{r}+U_{r}^{2}\chi _{m\mathbf{q}}\text{,}  \label{vcor}
\end{equation}%
where $\chi _{m\mathbf{q}}$ is the electronic susceptibility. The positive
constant $U_{r}$ in Eq.(\ref{vcor}) is just the direct first order Coulomb
repulsion suppressing the s-wave pairing, but having no impact on the d -
wave pairing.

The well known Kohn-Luttinger diagrams\cite{Kohn,Maier20} give in the
overdoped case, $x>x^{\ast }$, the following dynamic Matsubara
susceptibility:%
\begin{equation}
\chi _{m\mathbf{q}}=\frac{1}{N^{2}}\sum \nolimits_{\mathbf{p}}\frac{f_{F}%
\left[ E_{\mathbf{p+q}}\right] -f_{F}\left[ E_{\mathbf{p}}\right] }{i\omega
_{m}+E_{\mathbf{p}}-E_{\mathbf{p+q}}}\text{,}  \label{suss}
\end{equation}%
where $E_{\mathbf{p}}$ was defined in Eq.(\ref{EHFdef}). This is calculated
numerically for sufficiently large values of $N=256$ and harmonics $%
\left
\vert m\right \vert \leqslant 32$. In the lower row of Fig.5 the
static part, namely zero frequency is given $x=x^{opt}=0.16$ and $x=0.2$.
Similarly in the underdoped case, $x<x^{\ast }$, one calculates the same two
diagrams on the magnetic BZ, $0<k_{1}<\pi ,-\pi <k_{2}<\pi $, namely using
the GF of Eq.(\ref{AFcorr}). Since we are interested in the dynamic
susceptibility on the scale of the Cooper pairs, the full sublattice matrix
should be used. This is derived in Appendix B, where a rather bulky
expression, Eq.(\ref{B_chaires}) is given. It turns out that after
symmetrization it is not much different from the overdoped case
susceptibility as is shown in Fig. 5. The symmetrization of the
susceptibility matrix is made as in ref.\cite{Li19}). The zero frequency $%
\chi _{0,k_{x},k_{y}}^{sym}$ at $T=50K$ is plotted for $x=0.13$. The
dependence on temperature in the relevant range ($T<300K$) is very weak. One
observes that the evolution is smooth through the Lifshitz point $x^{opt}$.

The general feature of the Matsubara susceptibility distribution over the BZ
is that near the crystallographic $M$ point the susceptibility is large,
while near the $\Gamma $ point it is small. This is crucial for the d - wave
pairing. Note also the fine structure of the susceptibility: there are two
characteristic local maxima near point $M$, while the point itself is a
local minimum. The splitting is very small. \ In this paper we do not
consider possible fourfold symmetry breaking (or nematicity). This effective
electron - electron couplings will be used in the gap equation.

\subsection{Superconducting gap}

To complete the electronic effective action, one adds to Eqs.(\ref{Aee}) and
(\ref{vcor}) the electronic part, 
\begin{equation}
\mathcal{A}_{eff}=\frac{1}{T}\sum \nolimits_{n\mathbf{k}}\left \{ \psi _{n%
\mathbf{k}}^{\ast \sigma }G_{n\mathbf{k}}^{-1}\psi _{n\mathbf{k}}^{\sigma }+%
\frac{1}{2}n_{n\mathbf{k}}v_{n\mathbf{k}}n_{-n,-\mathbf{k}}\right \} ;
\label{actionpara}
\end{equation}%
where $v_{n\mathbf{q}}=v_{n\mathbf{q}}^{ph}+v_{n\mathbf{q}}^{cor}$, $G$ is
the (HF) Green's function and $v^{ph}$ and $v^{cor}$ are given by Eq.(\ref%
{Aee}) and Eq.(\ref{vcor}) respectively. The standard superconducting gap
equation is,

\begin{equation}
\Delta _{m\mathbf{k}}=-\frac{T}{N^{2}}\sum \nolimits_{n\mathbf{p}}\frac{%
v_{m-n,\mathbf{k-p}}}{G_{n\mathbf{p}}^{-1\ast }\Delta _{n\mathbf{p}}^{-1}G_{n%
\mathbf{p}}^{-1}+\Delta _{n\mathbf{p}}^{\ast }}\text{.}  \label{gapeqpara}
\end{equation}%
Here the (Matsubara) gap function is related to the anomalous GF, $%
\left
\langle \psi _{m\mathbf{k}}^{\sigma }\psi _{n\mathbf{p}}^{\rho
}\right
\rangle =\delta _{n+m}\delta _{\mathbf{k+p}}\varepsilon ^{\sigma
\rho }F_{m\mathbf{k}}$ ($\varepsilon ^{\sigma \rho }$ - the antisymmetric
tensor), by 
\begin{equation}
\Delta _{m\mathbf{k}}=\frac{T}{N^{2}}\sum \nolimits_{n\mathbf{p}}v_{m-n,%
\mathbf{k-p}}F_{n,\mathbf{p}}\text{.}  \label{deltasupdef}
\end{equation}%
The gap equation was solved numerically by iteration for $N=256$ and $64$
frequencies. It converges to the d - wave solution. An example of the gap
distribution over the BZ (for the optimal doping at $T=50K$) is given in
figure in Appendix C. The absolute value of the Matsubara gap function has a
maximum near the crystallographic $X$ point $\left( 0,\pi \right) $. This
value as function of doping and temperature is given in figure in Appendix C.

In an AF domain (considered to be larger than the Cooper pair) the fourfold
symmetry is broken. As a consequence one uses basis consisting of two
sublattices $I=A,B$ and the magnetic BZ defined in Appendix B. The
electronic effective action, in this basis takes a form\bigskip 
\begin{equation}
\mathcal{A}_{eff}=\frac{1}{T}\sum \nolimits_{n\mathbf{k}}\left \{ \psi _{n%
\mathbf{k}}^{\ast \sigma I}\left[ G_{n\mathbf{k}}^{-1\sigma }\right]
^{IJ}\psi _{n\mathbf{k}}^{\sigma J}+\frac{1}{2}n_{n\mathbf{k}}^{\sigma I}v_{n%
\mathbf{k}}^{\sigma \rho IJ}n_{-n,-\mathbf{k}}^{\rho J}\right \} \text{,}
\label{actionAF}
\end{equation}%
where $G$ is the (HF) Green's function is given in Eq.(\ref{AFcorr}) in
Appendix B. The symmetrized susceptibility in the underdoped cases of $%
x=0.02,0.05,0.13$ are given in Fig.5. One observes that the distribution is
continuously crosses over to the overdoped one via the (Lifshitz)
topological transition at optical doping.

The anomalous Green's function is also a $2\times 2$ matrix in sublattice
space. For singlet pairing one has: $\left \langle \psi _{n,\mathbf{k}%
}^{\sigma I}\psi _{-n,-\mathbf{k}}^{\rho J}\right \rangle =\varepsilon
^{\sigma \rho }F_{n\mathbf{k}}^{IJ}$. Assuming the up-down (singlet) pairing%
\cite{Rosen19}, see Appendix C,

\begin{equation}
\left[ \Delta _{n\mathbf{k}}\right] =%
\begin{array}{cc}
0 & \Delta _{n\mathbf{k}}^{\uparrow \downarrow } \\ 
\Delta _{n\mathbf{k}}^{\downarrow \uparrow } & 0%
\end{array}%
;\text{ \  \  \ }\Delta _{n\mathbf{k}}^{\uparrow \downarrow IJ}=\sum
\nolimits_{m\mathbf{p}}v_{n-m,\mathbf{k}-\mathbf{p}}^{\uparrow \downarrow
IJ}F^{\uparrow \downarrow IJ}\text{,}  \label{delAF}
\end{equation}%
the gap equation in matrix form becomes,%
\begin{equation}
\left[ \Delta _{n\mathbf{k}}^{\uparrow \downarrow }\right] =-\sum
\nolimits_{m\mathbf{p}}\left[ v_{n-m,\mathbf{k}-\mathbf{p}}\right] \ast
\left \{ \left[ G_{m\mathbf{p}}^{-1\downarrow }\right] ^{\dagger }\left[
\Delta _{m\mathbf{p}}^{\uparrow \downarrow }\right] ^{-1}\left[ G_{m\mathbf{p%
}}^{-1\uparrow }\right] +\left[ \Delta _{m\mathbf{p}}^{\uparrow \downarrow }%
\right] ^{\dagger }\right \} ^{-1}\text{,}  \label{GapeqAF}
\end{equation}%
and the same for $\Delta _{n\mathbf{k}}^{\downarrow \uparrow }$. The star
product denotes the matrix element multiplication. The iteration solution
for the same system size, as in the overdoped case, converges to the d -
wave solution for wide range of initial conditions. The results for various
temperatures are given in SM C, while critical temperatures with (without)
spin fluctuations are presented as black (green) point in Fig.4. In the
concluding section the results are qualitatively discussed.

The line of vanishing gap determines the $T_{c}$ values on the phase diagram
in Fig.4 (red squares). In the underdoped domain it comes short of the
parabolic experimental dependence\cite{accurate} (dashed curve). If one
neglects the magnon contribution, namely takes $v=v^{ph}$, the temperatures
are lower by 15-20\% (red circles).

\subsection{Isotope effect}

The influence of the oxygen isotope substitution, $^{16}O\rightarrow $ $%
^{18}O$ on superconductivity can be gauged by calculation of the change of
the (Matsubara) gap at a temperature below $T_{c}$. In Fig.6 we plot the The
deduced exponent, 
\begin{equation}
\alpha =\frac{18}{16}\log \frac{\Delta \left( ^{16}O\right) }{\Delta \left(
^{18}O\right) }\text{,}  \label{alfa}
\end{equation}%
at temperature $T=20K$.

The same exponent was estimated by measuring the $T_{c}$ isotope effect in
various hole doped cuprates\cite{Muller}, mostly in $YBa_{2}Cu_{3}O_{7-x}$
and $La_{2-x}Sr_{x}CuO_{4}$. Qualitatively the exponent is small in
overdoped and optimally doped materials, but increases at strongly overdoped
case. In $Ba_{2}Sr_{2}CaCu_{2}O_{7}$ the experimental results are scarce,
but order of magnitude is the same as in Fig. 6.

\begin{figure}[h]
\centering \includegraphics[width=10cm]{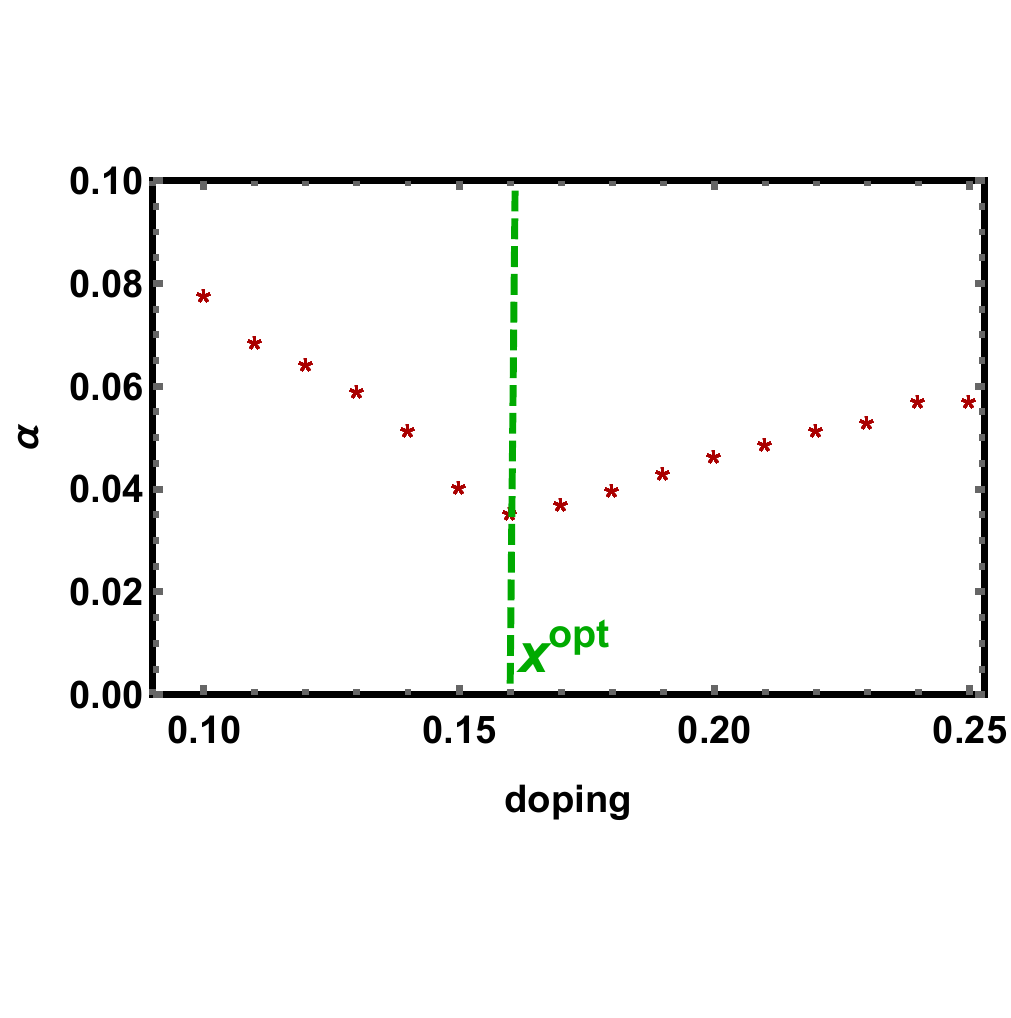}
\caption{Isotopic effect critical exponent versus doping}
\label{Fig,6}
\end{figure}
The isotope effect exponential is small, at optimal and overdoped systems,
however it slightly increases when the doping is reduced below optimal
(reaches $\alpha =0.08$ at $x=0.1$).

\subsection{Thermal fluctuations and the inter-layer tunneling}

In the bi-layer $Bi2212$ there are two types of tunneling. The first is a
rather strong tunneling between adjacent layers within the bi-layer is
estimated\cite{Kordyuk} to be $t_{\perp }^{\prime }=30-80meV$. It leads to
appearance of the secondary band mentioned in Section II. The second
tunneling amplitude is between the bi-layers in different cells. The 3D
dispersion relation is $\epsilon _{\mathbf{k,}k_{z}}=\epsilon _{\mathbf{k}%
}+\epsilon _{\mathbf{k}}^{\prime }+t_{\perp }\cos \left( k_{z}s\right) $,
where $\epsilon _{\mathbf{k}},\epsilon _{\mathbf{k}}^{\prime }$ are given in
Eq. (\ref{tight1}) and $s$\ - the inter bi-layers separation. The order of
magnitude is much smaller than $t_{\perp }^{\prime }$: $t_{\perp }=1-2meV$.

Superconductivity in a single $CuO$ bi-layer is of the 2D Kosterlitz
-Thouless type. The mean field critical temperature calculated above
slightly overestimates $T_{KT}$, where the modulus of the order parameter is
established: $T_{c}-T_{KT}\approx T_{c}Gi_{2D}$. Here $Gi_{2D}$ is the 2D
Ginzburg number, $Gi_{2D}=a^{2}/d\xi _{\parallel }$, $d$ is the thickness of
the CuO bi-layer and $\xi _{\parallel }$\ is the lateral coherence length%
\cite{gausspert}. Due to the tunnelings $t_{\perp }$ between bi - layers
makes the system 3D and the KT feature disappears.

The 2D/3D crossover occurs when $\xi _{\perp }\sim s$ where $\xi _{\perp }$
are the coherence length in $z$ - direction. Close to the critical
temperature $\xi _{\perp }\sim \hbar v_{z}/T_{c}\sqrt{1-T/T_{c}}$, where $%
\hbar v_{z}=\frac{\partial \epsilon }{\partial k_{z}}=t_{\perp }s$. It
determines the temperature range in which the superconductivity is
essentially 3D: $\left \vert 1-T/T_{c}\right \vert <\left( t_{\perp
}/T_{c}\right) ^{2}$ $\sim $ $0.02$.

\section{Discussion and conclusions.}

\begin{figure}[h]
\centering \includegraphics[width=12cm]{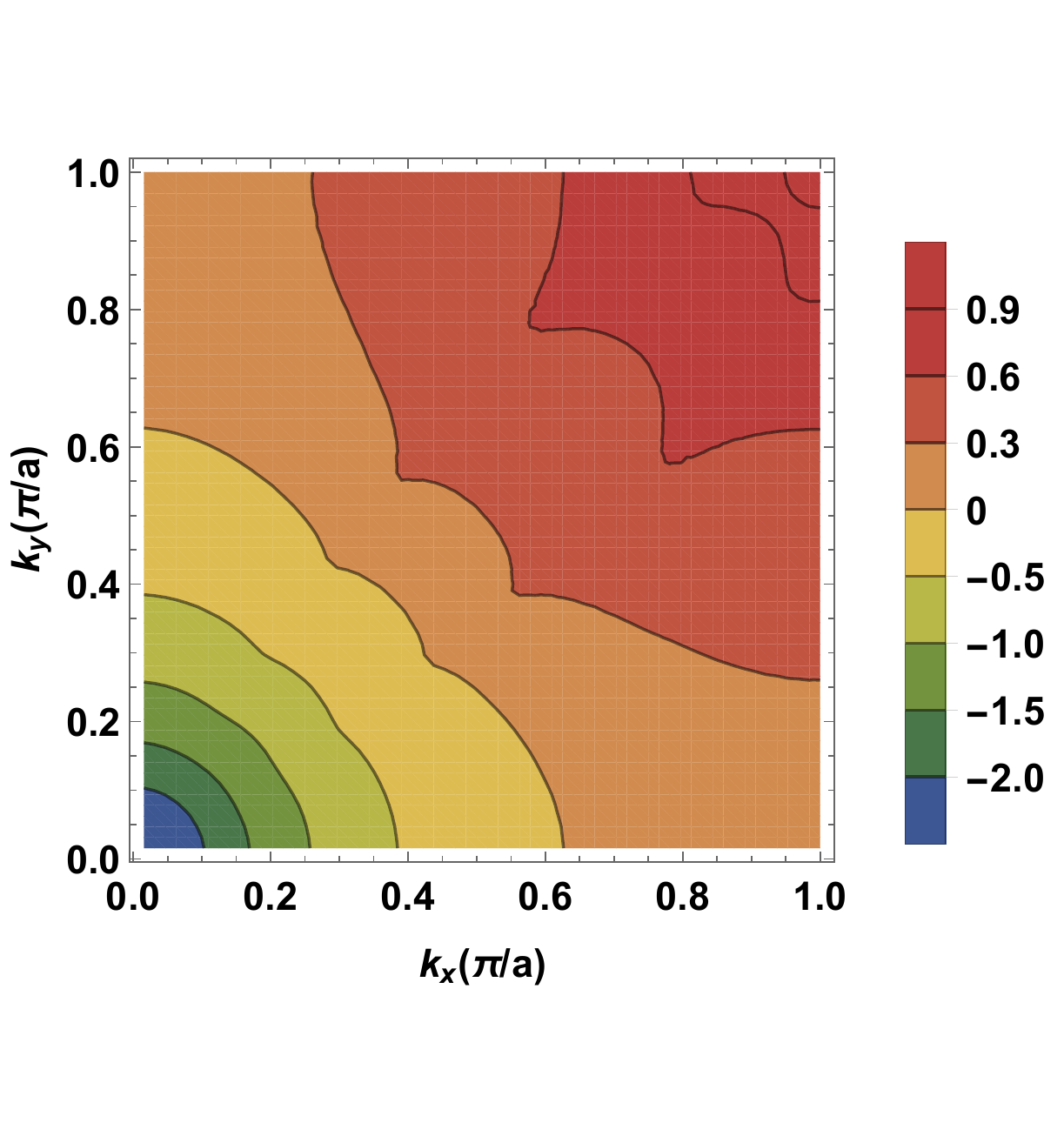}
\caption{The overall potential including both the phonon central dip \ at $%
\Gamma $ and the correlation peak at $M$.. Note that the dip is larger than
the peak leading to dominance of the phonon channel.}
\end{figure}

\subsection{Summary}

Theory of superconductivity of high $T_{c}$ cuprates based on the dominant
ALLP pairing mechanism was proposed. It is comprehensive in a sense that the
whole doping range is considered including anomalous normal state properties
of cuprates like $Bi_{2}Sr_{2}CaCu_{2}O_{8+x}$. To demonstrate the basic
principles we limited ourselves in this paper to a simple sufficiently
generic model: the pseudogap physics of 2DEG in the $CuO$ planes is
described by the fourfold symmetric $t-t^{\prime }$ single band Hubbard
model with on site repulsion energy $U$ of moderate strength. Doping is
controlled by the chemical potential.

The results are following. The most important for the pairing mode for $%
Bi2212$ is found to be the optical longitudinal\ lateral (within the $SrO$
plane) mode at $45meV$, mostly due to vibration of apical oxygen atoms. The
dimensionless electron - electron attraction exhibits an exponential forward
scattering peak and is estimated to have the strength of $\lambda \sim 0.6$.
When parameters of the effective one band $\ t-t^{\prime }$ model of \ 2DEG
were fixed at $t^{\prime }\sim -0.17t$ and $U\sim 6t$, $t=0.3eV$, the mean
field $T^{\ast }\,$line, green curve in phase diagram, fig.3, become a
crossover between short range AF pseudogap phase and the paramagnetic one.
The quasi - particle spectrum undergoes a topological (Lifshitz) transition.
The closed Fermi surface above the $T^{\ast }$ line disintegrates into four
Fermi arcs below it, see Fig. 4.

Renormalization of the electron Green's function due to phonons allows
calculation of the quasi - particle properties. Location of kink in
dispersion relation including the observed isotope ($^{16}O\rightarrow
^{18}O $) dependence, see Fig. 5. Since the electron - phonon coupling $%
\lambda $ is moderate, weak coupling dynamic Eliashberg approach is
applicable to calculate the gap function and critical temperature $T_{c}$.
One has to go beyond the BCS approximation due to important dependence of
the phonon mediated pairing on frequency. Both phonon and spin fluctuation
pairing are accounted for over the full doping range. It is found that the
critical temperatures above $90K$ at optimal doping can be reached, see Fig.
4. The dominant "glue" responsible for the d - wave pairing turns out to be
the phonon mode rather than spin fluctuations, although the later enhances
superconductivity by about 15-20\%. Comparison of the doping dependence of $%
T_{c}$with experimental\cite{accurate}is qualitatively fair, although .
underdoped are slightly underestimated, while strongly overdoped
overestimated. The isotope ($^{16}O\rightarrow ^{18}O$\ substitution) effect
is small at optimal doping but increases towards both the underdoped and the
overdoped regions, see Fig.6. This is consistent with observations\cite%
{BSCCOisotope}.

\subsection{Qualitative picture of the d-wave superconductivity}

Let is now make an argument qualitatively describing the d - wave pairing by
ALLP and its coexistence with spin fluctuations or other pairing "glue".
Generally the pairing potential $v$ should have sufficiently large
dependence on momentum over BZ. The overall pairing potential, Eq.(\ref%
{actionpara}) is a sum of the phonon and the spin fluctuations
contributions. \ The ALLP's forward scattering peak presents itself as a
large dip of the potential at the $\Gamma $ point, see Fig. 7, due to
attractive nature of the EPI. In contrast the spin susceptibility peak of
Fig. 4 causes a smaller maximum \ref{xdef} at the $M$ point (corner of
Brillouin zone), since the interaction is repulsive. Both regions of the BZ
contribute to d-wave superconductivity and fortunately do no interfere with
each other. Indeed the phonon peak decreases exponentially to just 10\% at
distance $k^{ph}=\frac{1}{d_{a}}=\frac{2\pi }{3a}$, where $d_{a}$ is the
vertical distance of the $CuO$ layer from the $SrO$ layer, see Fig. 1. The
susceptibility becomes negligible at distance $\frac{\pi }{3a}$ from $M$,
see Fig.4. Hence the BZ is effectively utilized.

To summarize, two features turned out to be sufficient for robust apical
phonon d - wave pairing. The first is the rhombic shape of the Fermi
surface. The second is the exponential FSP of the apical lateral phonon
optical mode and, to a lesser degree, constructive cooperation with the spin
fluctuation channel. The s-wave solution of the gap equation sometimes
competes with the d - wave that appears only when the fourfold anisotropy of
the Fermi surface is sufficiently pronounced. In these cases the central
peak favors d-wave over the s-wave due to two reasons. First, the s-wave
pairing due to the apical phonons is generally weaker than the $CuO$ plane
phonons since unlike in BCS large momentum $q$ contributions are suppressed.
Second, while the s-wave channel is suppressed by direct Coulomb repulsion,
the d-wave \textit{is not} (the quasi - local Coulomb repulsion drops out of
the gap equation for the d-wave). We have explicitly compared energies and
found that the s-wave loses in the range presented.

\subsection{Concluding remarks}

Restriction of the description of the electron gas to one band Hubbard model
with just two parameters $t,t^{\prime }$ for nearest neighbor and next to
nearest neighbor hopping obviously makes the model less realistic to
quantitatively describe real materials like $Bi2212$. These typically
require either a three band much more complicated model or an effective one
band model with more distant hopping terms like $t^{\prime \prime }$. In
addition the tunneling between the conducting $CuO$ planes via a metallic
layer and the nematicity (deviations from the fourfold symmetry) should be
added. These lead to a characteristic splitting of the spectrum\cite{Kordyuk}%
. This is left for future work. Of course the phenomena broadly termed "
unusual normal and superconducting properties of high $T_{c}$ cuprates "
contains many more features. In this paper we have emphasized ones that are
directly linked to the phonon exchange.

Experimentally the main claim of the paper, namely that the "glue" that
creates d - wave pairing is the phonon exchange of a very specific nature,
the apical oxygen's (that is one belonging to an insulating layer, $SrO$,
adjacent to the conducting $CuO$ layer) lateral vibrations, can be further
directly strengthened or falsified by suppression such vibrations as in refs%
\cite{smokinggun}\cite{DavisBalatsky} or actively focus on these modes and
their coupling. Since one or to unit cell perovskite were recently fabricated%
\cite{XueBSCCO}\cite{accurate}\cite{Kim2UC} perhaps apical oxygen atoms can
be distinguished from the rest. An alternative route is to look for
secondary effects of this coupling on normal state properties, some
calculated in the present paper. The phonons induce modifications in normal
state like modification of dispersion relation on transport beyond the
"strange metal" resistivity behavior. The modification can be isolated by
isotope substitution. Superconducting properties due to this particular
mechanisms in addition to $T_{c}$ and order parameter studied, are also
sensitive to the isotope substitution. An example is magnetization curves%
\cite{isotopemag} that simply depend on $T_{c}$ (via Ginzburg - Landau
description\cite{gausspert}).

\textit{Acknowledgements. }

We are grateful Prof. D. Li, H.C. Kao, T. X. Ma, L. L.Wang, Y. Guo, J.Y. Lin
and Y. Yeshurun for helpful discussions. Work of B.R. was supported by NSC
of R.O.C. Grants No. 101-2112-M-009-014-MY3.

\appendix{}

\section{A. The apical oxygen lateral vibration modes and their coupling to
holes in the CuO plane.}

\subsection{The apical oxygen lateral branches}

The approximate method of determining the relevant vibration modes is the
same as previously used for the $FeSe$ on STO superconductor, see details in
Appendix A of ref. \cite{Rosen19}. The chart on the right in Fig. A1 is a
view from above with sphere radii corresponding to the repulsive Born -
Meyer potential ranges given in Table I. Unit cell including both the
metallic layer and the substrate is marked by the black frame in Fig. 8.
Dynamic degrees of freedom are the $O$ atoms in the $SrO$ layer, see Fig.8.

\begin{figure}[h]
\centering \includegraphics[width=16cm]{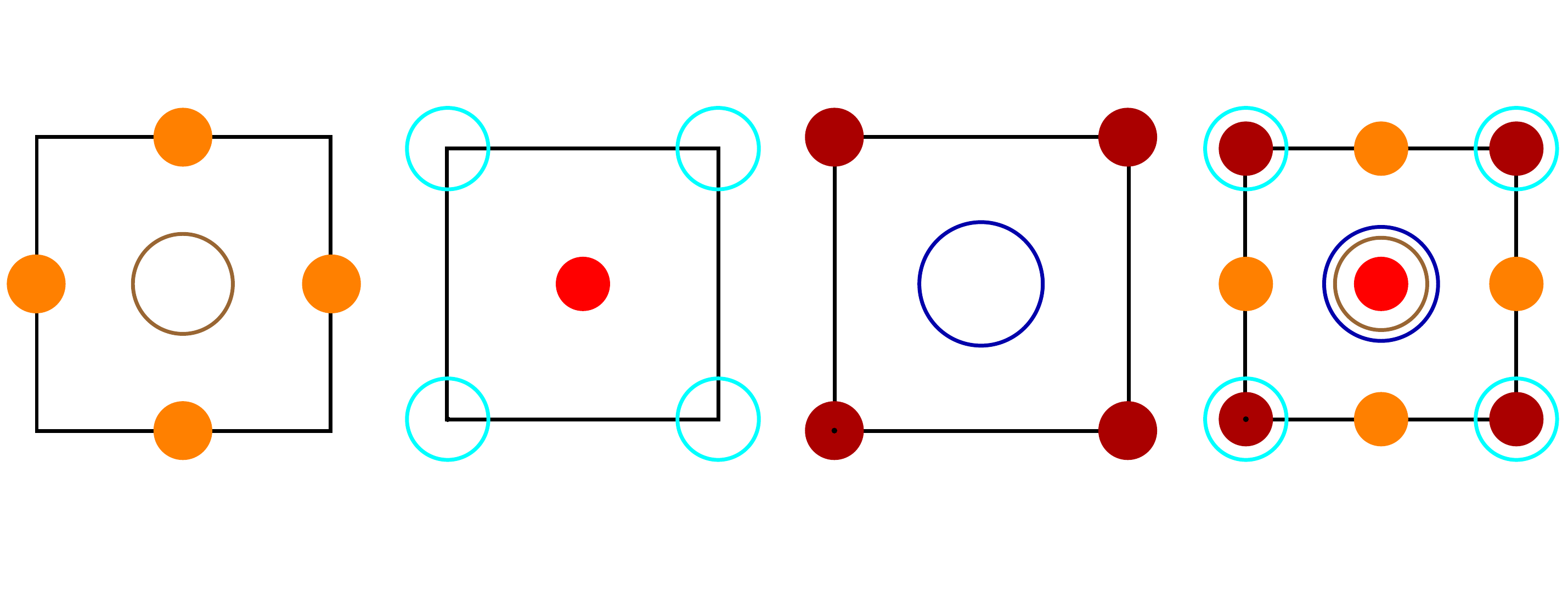}
\caption{ Atomic lateral positions of the three layers a. the 2DEG layer
consisting of $Cu$ at $\mathbf{R}^{Cu}=\left( 0,0,z^{Cu}\right) $ and two $%
O_{1}$ atoms at $\mathbf{R}^{Ox}=\left( a,a/2,z^{Cu}\right) $ and $\mathbf{R}%
^{Oy}=\left( a/2,a,z^{Cu}\right) $. b. the apical phonon layer containing
the $Sr$ at origin $\mathbf{R}^{Sr}=\left( 0,0,0\right) $ and $\mathbf{R}%
^{Oy}=\left( a/2,a,z^{Cu}\right) $ and $O_{2}$ at $\mathbf{R=}\left(
a/2,a/2,0\right) $. c. the third layer: $Bi$ at $\mathbf{R}^{Bi}=\left(
a/2,a/2,z^{Bi}\right) $ and $O_{3}$ at $\mathbf{R}^{O3}=\left(
0,0,z^{Bi}\right) $. d. The top view: all the three layer's projections are
superimposed.}
\end{figure}

Hamiltonian for these degrees of freedom is:%
\begin{equation}
H_{ph}=K_{ph}+W\text{.}  \label{A_Hph}
\end{equation}%
Here kinetic energy is%
\begin{equation}
K_{ph}=\frac{M}{2}\sum \nolimits_{\mathbf{n}}\left( \frac{d}{dt}\mathbf{u}_{%
\mathbf{n}}\right) ^{2}\text{,}  \label{A_Kph}
\end{equation}%
while the potential energy part $W$ consists of interatomic Born - Meyer
potentials defined in Eq.(\ref{interatomic}). Only interactions of the
"dynamic" oxygen atoms in the $SrO$ with neighboring $BiO$ below and $%
CuO_{2} $ above are taken into account:%
\begin{eqnarray}
W &=&\frac{1}{2}\sum \nolimits_{\mathbf{n,m}}\left \{ 
\begin{array}{c}
v^{SrO}\left[ \mathbf{R}^{Sr}+\mathbf{r}_{\mathbf{n}}-\mathbf{r}_{\mathbf{m}%
}-\mathbf{u}_{\mathbf{m}}\right] +v^{CuO}\left[ \mathbf{R}^{Cu}+\mathbf{r}_{%
\mathbf{n}}-\mathbf{r}_{\mathbf{m}}-\mathbf{u}_{\mathbf{m}}\right] \\ 
+v^{OO}\left[ \mathbf{R}^{Ox}+\mathbf{r}_{\mathbf{n}}-\mathbf{r}_{\mathbf{m}%
}-\mathbf{u}_{\mathbf{m}}\right] +v^{OO}\left[ \mathbf{R}^{Oy}+\mathbf{r}_{%
\mathbf{n}}-\mathbf{r}_{\mathbf{m}}-\mathbf{u}_{\mathbf{m}}\right] \\ 
+v^{BiO}\left[ \mathbf{R}^{Bi}+\mathbf{r}_{\mathbf{n}}-\mathbf{r}_{\mathbf{m}%
}-\mathbf{u}_{\mathbf{m}}\right] +v^{OO}\left[ \mathbf{R}^{O3}+\mathbf{r}_{%
\mathbf{n}}-\mathbf{r}_{\mathbf{m}}-\mathbf{u}_{\mathbf{m}}\right]%
\end{array}%
\right \}  \label{A_potW} \\
&&+\frac{1}{2}\sum \nolimits_{\mathbf{n\not=m}}v^{OO}\left[ \mathbf{r}_{%
\mathbf{n}}-\mathbf{r}_{\mathbf{m}}+\mathbf{u}_{\mathbf{n}}-\mathbf{u}_{%
\mathbf{m}}\right] \text{.}  \notag
\end{eqnarray}%
Here the lateral apical oxygen positions are 
\begin{equation}
\text{ }\mathbf{r}_{\mathbf{m}}=a\left( m_{1},m_{2}\right) ,  \label{A_R}
\end{equation}%
while positions of the heavy $Sr,Cu,Bi$ , see Figs. 8, are 
\begin{eqnarray}
\mathbf{R}^{Sr} &=&a\left( \frac{1}{2},\frac{1}{2},0\right) ;  \label{A-pos}
\\
\mathbf{R}^{Cu} &=&z\mathbf{_{Cu}\widehat{\mathbf{z}}};  \notag \\
\mathbf{R}^{Bi} &=&z\mathbf{_{Bi}\widehat{\mathbf{z}}}\text{.}  \notag
\end{eqnarray}%
The inter-layer spacings are given in Table I and $\mathbf{\widehat{\mathbf{z%
}}\equiv }\left( 0,0,1\right) $. The positions of the two oxygen atoms of
the $CuO_{2}$ layer and that in the $BiO$ layer are:%
\begin{equation}
\mathbf{R}^{Ox}=\left( 0,a/2,z\mathbf{_{Cu}}\right) ;\text{ \ }\mathbf{R}%
^{Oy}=\left( a/2,0,z\mathbf{_{Cu}}\right) ;\text{ }\mathbf{R}^{O3}=\left(
0,0,z\mathbf{_{Bi}}\right) .  \label{A_posO}
\end{equation}%
Consequently the dominant lateral displacements, $u_{\mathbf{m}}^{\alpha }$, 
$\alpha =x,y$, are of the apical oxygen atoms.

Vibrations of heavy atoms and even oxygen in other planes are not expected
to be significant due to their mass or distance from the $SrO$ layer oxygen
atoms. Some effects of those vibrations can be accounted for by the
effective oxygen mass, while more remote layers above and below the
important layer were checked to be negligible.

Harmonic approximation consists of expansion around a stable minimum of the
energy. Expressions for the derivatives are given in ref.\cite{Rosen19}.
This leads to the following expression for the dynamic matrix

\begin{eqnarray}
D_{\mathbf{k}}^{\alpha \beta } &=&\sum \nolimits_{\mathbf{n}}\left[ v^{CuO}%
\right] _{\alpha \beta }^{\prime \prime }[\mathbf{R}^{Cu}\mathbf{+r}_{%
\mathbf{n}}]+\left[ v^{OO}\right] _{\alpha \beta }^{\prime \prime }[\mathbf{R%
}^{Ox}+\mathbf{r}_{\mathbf{n}}]  \notag \\
&&+\left[ v^{OO}\right] _{\alpha \beta }^{\prime \prime }[\mathbf{R}^{Oy}+%
\mathbf{r}_{\mathbf{n}}]+\left[ v^{SrO}\right] _{\alpha \beta }^{\prime
\prime }[\mathbf{R}^{Sr}+\mathbf{r}_{\mathbf{n}}]+\left[ v^{BiO}\right]
_{\alpha \beta }^{\prime \prime }[\mathbf{R}^{Bi}\mathbf{+r}_{\mathbf{n}}] 
\notag \\
&&+\left[ v^{BiO}\right] _{\alpha \beta }^{\prime \prime }[\mathbf{R}^{O3}%
\mathbf{+r}_{\mathbf{n}}]+\left( 1-\exp \left[ -i\mathbf{k\cdot r}_{\mathbf{n%
}}\right] \right) \left[ v^{OO}\right] _{\alpha \beta }^{\prime \prime }%
\left[ \mathbf{r}_{\mathbf{n}}\right] \text{.}  \label{A_Dk}
\end{eqnarray}%
These matrix elements determine the frequencies (eigenvalues) for the two
polarizations presented in Fig. 2.

\subsection{The electron - phonon matrix elements}

The microscopic derivation of the electron - phonon coupling of the holes
residing in the $CuO$ plane should in principle start at least from an
effective three band (Emery) model of cuprate\cite{oneband}. It is often
reduced to the two band\cite{ZhangRice} model consisting of the Zhang - Rice
singlet state, a symmetric combination of the (in plane) $2p_{x}$ and $%
2p_{y} $\ $O$ orbitals, and the $3d_{x^{2}-y^{2}}$ $Cu$ orbitals. Let us
assume a simplified picture case that the hole's wave function is
concentrated on two oxygen positions within the unit cell. Concentrating on
one unit cell, drawn in Fig. 8a (left) as a black square, the 2D density is:

\begin{eqnarray}
\left \vert \varphi \left( \mathbf{r}\right) \right \vert ^{2} &=&\frac{1}{4}%
\left \{ 
\begin{array}{c}
\delta \left( \mathbf{r}+a\left( 1/2,0\right) \right) +\delta \left( \mathbf{%
r}-a\left( 1/2,0\right) \right) \\ 
+\delta \left( \mathbf{r}+a\left( 0,1/2\right) \right) +\delta \left( 
\mathbf{r}-a\left( 0,-1/2\right) \right)%
\end{array}%
\right \} ;  \label{wave} \\
n_{\mathbf{l}}\left( \mathbf{r}\right) &=&\frac{1}{N^{2}}\left \vert \varphi
\left( \mathbf{r-r}_{\mathbf{l}}\right) \right \vert ^{2}\text{.}
\end{eqnarray}%
Extent of the density distribution in the $z$ direction is neglected.

The electron - ion electrostatic energy is,

\begin{equation}
H_{ei}=-e\int_{\mathbf{r}}\Phi \left( \mathbf{r}\right) n\left( \mathbf{r}%
\right) \text{,}  \label{3_localized}
\end{equation}%
where the potential was given by Eq.(\ref{3_pot}) and the electron density
(due to the localization of the wave functions independent of the electron
quasi - momentum $\mathbf{k}$, see ref.\cite{Davidov}), $n\left( \mathbf{r}%
\right) \varpropto \sum \nolimits_{\mathbf{l}}n_{\mathbf{l}}\left( \mathbf{r}%
\right) $. Substituting and expanding to first order in the oxygen
vibrations, one obtains:

\begin{equation}
H_{ei}=-Ze^{2}\sum \nolimits_{\mathbf{l},\mathbf{m}}\int_{\mathbf{r}}\frac{%
\left( \mathbf{r}-\mathbf{r}_{\mathbf{m}}\right) \cdot \mathbf{u}_{\mathbf{m}%
}}{\left( \left( \mathbf{r}-\mathbf{r}_{\mathbf{m}}\right)
^{2}+d_{a}^{2}\right) ^{3/2}}n_{\mathbf{l}}\left( \mathbf{r}\right) \text{.}
\end{equation}%
Fourier transforming $u$ this takes a form:%
\begin{equation}
H_{ei}=Ze^{2}\sum \nolimits_{\mathbf{q}}n_{-\mathbf{q}}g_{\mathbf{q}%
}^{\alpha }\ u_{\mathbf{q}}^{\alpha }\text{,}
\end{equation}%
where $n_{\mathbf{q}}=\int_{\mathbf{r}}e^{i\frac{2\pi }{N}\mathbf{q\cdot r}%
}n_{\mathbf{l}}\left( \mathbf{r}\right) $ and with the matrix element

\begin{equation}
g_{\mathbf{q}}^{\alpha }=\frac{1}{2}\left( \cos \frac{aq_{x}}{2}+\cos \frac{%
aq_{y}}{2}\right) \overline{g}_{\mathbf{q}}^{\alpha }\text{.}
\end{equation}%
The "local" matrix element function,%
\begin{equation}
\overline{g}_{\mathbf{q}}^{\alpha }=\sum \nolimits_{\mathbf{l}}e^{i\frac{%
2\pi }{N}\mathbf{q}\cdot \mathbf{l}}\frac{\mathbf{r}_{\mathbf{l}}^{\alpha }}{%
\left( \mathbf{r}_{\mathbf{l}}^{2}+d_{a}^{2}\right) ^{3/2}}\text{,}
\end{equation}%
would be obtained if the hole is localized right above the apical oxygen.
The summation in a very good approximation over the BZ can be replaced by
integration with the result given in Eq.(\ref{A}).

\begin{figure}[h]
\centering \includegraphics[width=10cm]{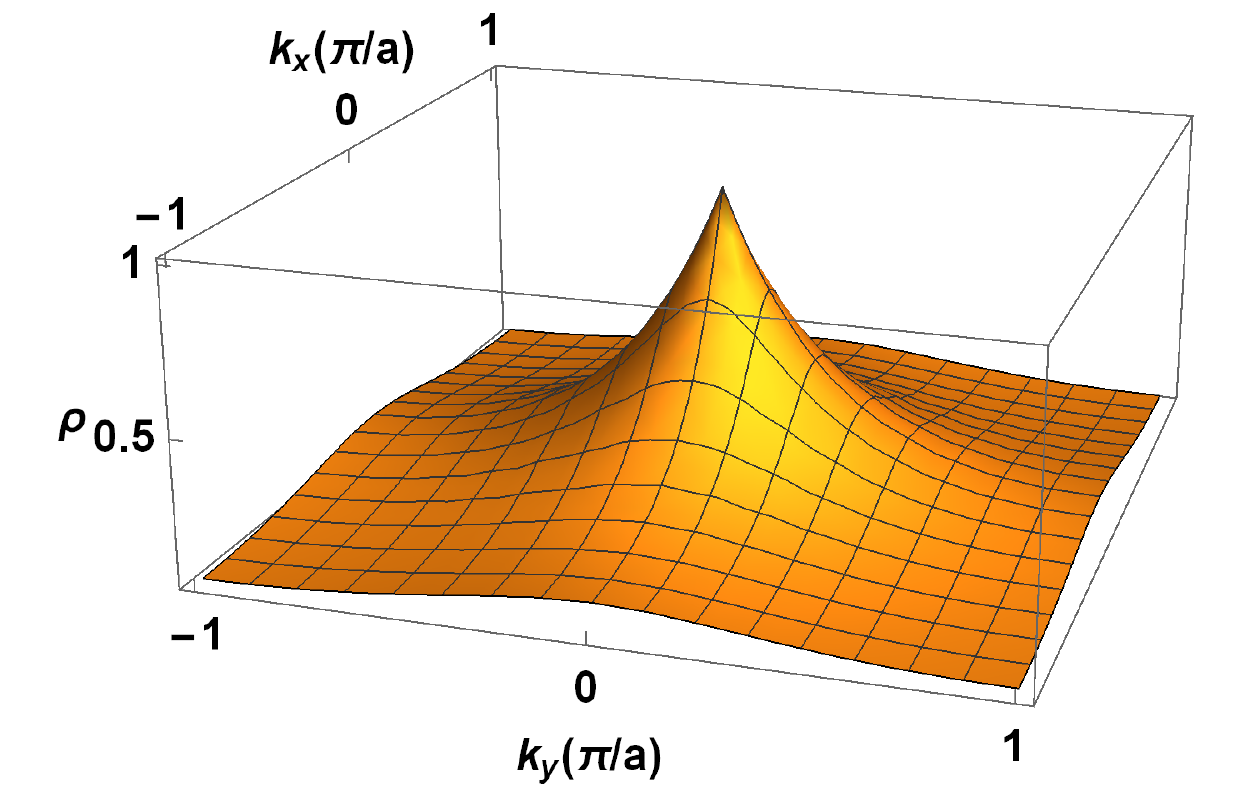}
\caption{Square of the matrix element of the electron - phonon coupling.
Decreases exponentially as function of quasi - momentum momentum away from
the $\Gamma $ point. The forward scattering peak region occupies a
significant portion of the Brillouin zone.}
\end{figure}

\begin{figure}[h]
\centering \includegraphics[width=10cm]{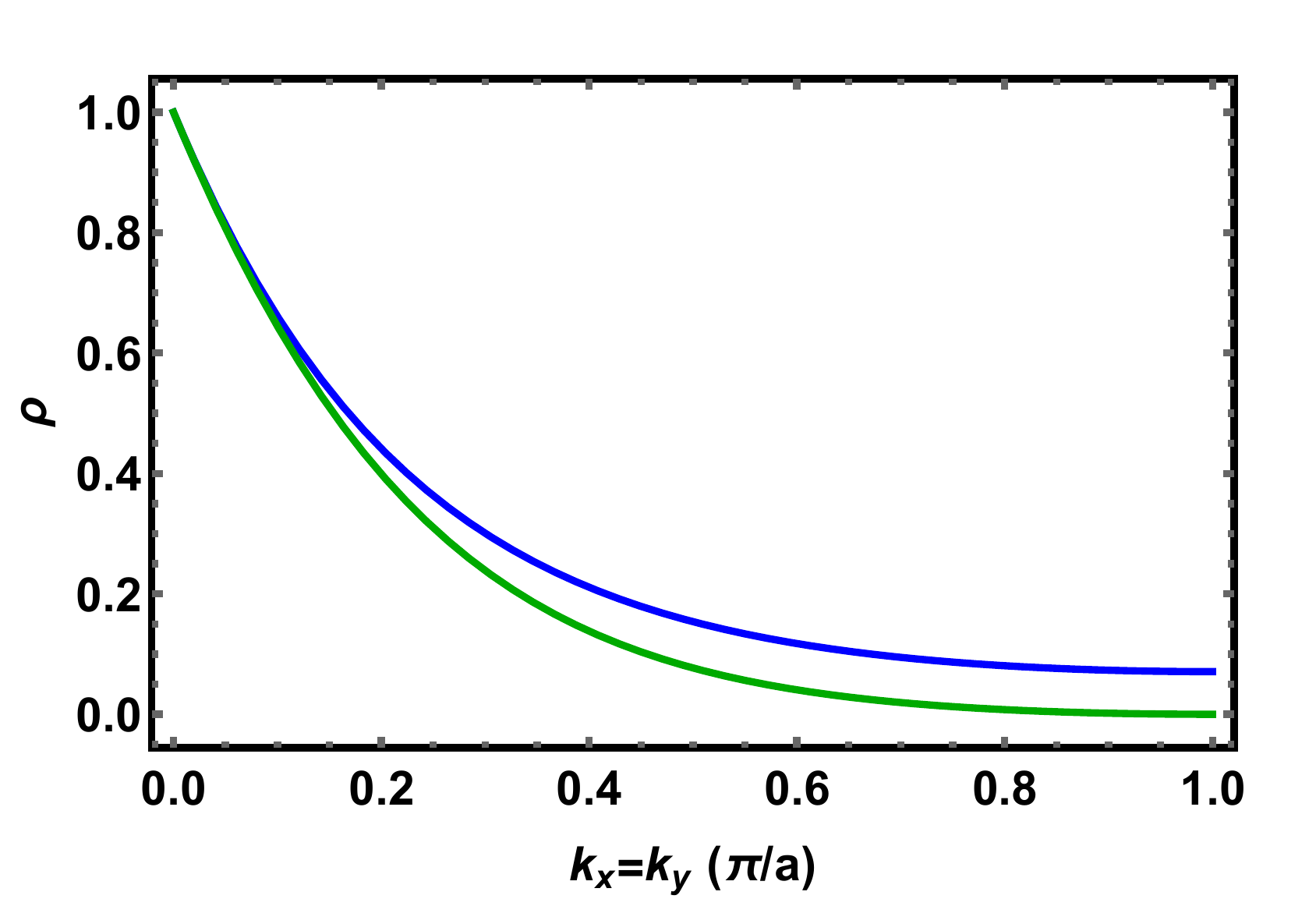}
\caption{Square of the matrix element of the electron - phonon coupling.
Decreases exponentially as function of quasi - momentum momentum away from
the $\Gamma $ point. The forward scattering peak region occupies a
significant portion of the Brillouin zone.}
\end{figure}

\section{B. Normal state properties}

\bigskip

\subsection{Coupling renormalization and the effective HF in overdoped and
underdoped regime.}

\bigskip

The HF theory of the $t-t^{\prime }$ model has been thoroughly investigated
over the years\cite{LinHQ}\cite{Wallin}. The spin rotation $SU\left(
2\right) $ symmetry in anti - ferromagnet is broken down to its $U\left(
1\right) $ subgroup. The on site magnetization, $M=\frac{1}{2}\left(
n^{A\downarrow }-n^{A\uparrow }\right) =\frac{1}{2}\left( n^{B\uparrow
}-n^{B\downarrow }\right) $, is considered to be oriented along the spin
space $z$ axis. The lattice translation symmetry consequently is reduced to
a smaller one on two sublattices $I=A,B$. The sublattice $A$ consists of odd 
$\left( i_{x}+i_{y}\right) $ sites, while $B$ contains even $\left(
i_{x}+i_{y}\right) $ sites. Position within the sublattices can be specified
by integers $i_{1}=1,...N/2\equiv N^{\prime }$ and $i_{2}=1,..N$, namely $%
c_{i_{1},i_{2}}^{A}=c_{2i_{1}-1+i_{2},i_{2}}$ and $%
c_{i_{1},i_{2}}^{B}=c_{2i_{1}+i_{2},i_{2}}$.

Hamiltonian in the magnetic quasi - momentum $\mathbf{k}$ space becomes
(integer momenta) is,

\begin{equation}
K=\sum \nolimits_{k_{1}k_{2}}\left \{ -\left( c_{\mathbf{k}}^{A\dagger }h_{%
\mathbf{k}}^{\ast }a_{\mathbf{k}}^{B}+h.c.\right) +c_{\mathbf{k}}^{I\dagger
}\left( \varepsilon _{\mathbf{k}}^{\prime }-\mu \right) c_{\mathbf{k}%
}^{I}\right \} \text{,}  \label{HubAF}
\end{equation}%
where

\begin{eqnarray}
h_{\mathbf{k}} &=&t\left \{ 1+\exp \left[ \frac{2\pi i}{N}\left(
2k_{1}-k_{2}\right) \right] +\exp \left[ \frac{2\pi i}{N^{\prime }}k_{1}%
\right] +\exp \left[ \frac{2\pi i}{N}k_{2}\right] \right \} ;  \label{hdef}
\\
\varepsilon _{\mathbf{k}}^{\prime } &=&-4t^{\prime }\cos \left[ \frac{2\pi }{%
N}k_{1}\right] \cos \left[ \frac{2\pi }{N}\left( k_{1}-k_{2}\right) \right] 
\text{.}  \notag
\end{eqnarray}%
The HF equations takes a form (using $n^{A\uparrow }\equiv
n_{1},n^{A\downarrow }=n_{2}$ electron densities on each site, no charge
density wave appear in the model considered), 
\begin{equation*}
n_{1}=F\left[ n_{1},n_{2}\right] ;\  \ n_{2}=F\left[ n_{2},n_{1}\right] ,
\end{equation*}
where the function $F$ is defined by

\begin{equation}
F\left[ n_{1},n_{2}\right] =\frac{1}{NN^{\prime }}\sum \nolimits_{\mathbf{k}%
}\left \{ f_{F}\left[ E_{\mathbf{k}}^{-}\right] -\frac{\Delta _{pg}+x_{%
\mathbf{k}}}{4x_{\mathbf{k}}}\left( \tanh \left[ \frac{E_{\mathbf{k}}^{+}}{2T%
}\right] -\tanh \left[ \frac{E_{\mathbf{k}}^{-}}{2T}\right] \right) \right
\} \text{.}  \label{Ffunction}
\end{equation}%
Here $\Delta _{pg}\equiv U_{r}M$ is the pseudogap energy and $f_{F}\left(
\varepsilon \right) \equiv \left( \exp \left[ \varepsilon /T\right]
+1\right) ^{-1}$ is the Fermi - Dirac distribution. The new quasi - particle
(hole in our case) spectrum consists of two branches%
\begin{equation}
E_{\mathbf{k}}^{\pm }=\varepsilon _{\mathbf{k}}^{\prime }-\mu +U_{r}\frac{%
n_{1}+n_{2}}{2}\pm x_{\mathbf{k}}\text{.}  \label{Edef}
\end{equation}%
and%
\begin{equation}
x_{\mathbf{k}}^{2}\equiv \Delta _{pg}^{2}+\left \vert h_{\mathbf{k}}\right
\vert ^{2}.  \label{xdef}
\end{equation}

The fact that the transition is second order is verified by the fitting of
the pseudogap curves near $T^{\ast }$ in Fig. 3 by a power $\Delta
_{pg}\propto \left( x-x^{\ast }\right) ^{\nu }$, with mean field critical
exponent $\nu =1/2$. It simultaneously satisfied the criticality condition
(where $n_{1}=n_{2}$):%
\begin{equation}
1=\frac{1}{NN^{\prime }}\sum \nolimits_{\mathbf{k}}\frac{U}{4x_{\mathbf{k}}}%
\left( \tanh \left[ \frac{E_{\mathbf{k}}^{+}}{2T^{\ast }}\right] -\tanh %
\left[ \frac{E_{\mathbf{k}}^{-}}{2T^{\ast }}\right] \right) \text{.}
\end{equation}

There is no experimental consensus on the shape of this line at small
temperatures\cite{NMR}\cite{AF}, while order of magnitude is consistent with
tunneling experiments\cite{Tstar}. In our model the low temperature segment, 
$T<T_{c},$\ of the line exhibits a weak first order transition with small
latent heat.

\subsection{Underdoped}

In 2D the Mermin - Wagner theorem \cite{Chaikin} states that fluctuations
for systems that have a continuous symmetry are strong enough to destroy
long range order at any nonzero temperature. The order parameter locally
exists, but averages out due to incoherence of its \textquotedblleft phase"
over the sample. A more rigorous approach would be to divide the degrees of
freedom into two scales, large distance correlations, and short distance
correlations. It can be performed for certain bosonic models using
renormalization group ideas, especially when the Berezinskii - Kosterlitz -
Thouless type transition is involved. However such an approach is
complicated in fermionic models in which order parameter is quadratic in
fermionic operators\cite{Metzner19}. A much simpler symmetrization approach
that does not involve the explicit separation of scales was proposed in ref.%
\cite{Li19}. It was demonstrated by comparing with determinantal Monte Carlo
simulations and for small sizes to exact diagonalization that he
symmetrization therefore qualitatively takes into account the largest
available scale by \textquotedblleft averaging over" the global symmetry
group and agrees to within 5\% with exact and MC results. We start with
symmetrization of the HF Green function (GF). For (conserved) spin
projection $\sigma $ the GF on magnetic BZ is a $2\times 2$ sublattice
matrix,

\begin{equation}
G_{mk_{1}k_{2}}^{\sigma }=\frac{1}{x_{\mathbf{k}}^{2}-\left( -i\omega
_{m}+E_{\mathbf{k}}^{\prime }\right) ^{2}}%
\begin{pmatrix}
-i\omega _{m}+E_{\mathbf{k}}^{\prime }-\left( -1\right) ^{\sigma }\Delta
_{pg} & h_{\mathbf{k}}^{\ast } \\ 
h_{\mathbf{k}} & -i\omega _{m}+E_{\mathbf{k}}^{\prime }+\left( -1\right)
^{\sigma }\Delta _{pg}%
\end{pmatrix}%
\text{,}  \label{AFcorr}
\end{equation}%
where $E_{\mathbf{k}}^{\prime }=\varepsilon _{\mathbf{k}}^{\prime }+\frac{%
U_{r}}{2}n-\mu $ and $\sigma =0$ for $\uparrow $ and $1$ for $\downarrow $.

\subsection{Symmetrization \ }

The relation between the matrix on magnetic Brillouin zone and the
symmetrized Matsubara Green's function on the whole BZ (nonmagnetic, since
the symmetry is restored), $-\pi /a<k_{x},k_{y}$ $\leq \pi /a$ is\cite{Li19},%
\begin{equation}
G_{mk_{x}k_{y}}^{sym}=\frac{1}{4}\sum \nolimits_{\sigma }\left(
G_{m,k_{x},k_{x}+k_{y}}^{\sigma
AA}+e^{ik_{x}a}G_{m,k_{x},k_{x}+k_{y}}^{\sigma
AB}+e^{-ik_{x}a}G_{m,k_{x},k_{x}+k_{y}}^{\sigma
BA}+G_{m,k_{x},k_{x}+k_{y}}^{\sigma BB}\right) \text{.}  \label{relation}
\end{equation}%
Here $G^{IJ}$ are elements of the matrix of Eq.(\ref{AFcorr}). As a result
the Green's function (after analytic continuation) is,%
\begin{eqnarray}
G^{sym}\left( \omega ,\mathbf{k}\right) &=&\frac{1}{2}\left( \frac{Z_{%
\mathbf{k}}^{+}}{\omega +i\eta +E_{\mathbf{k}}^{+}}+\frac{Z_{\mathbf{k}}^{-}%
}{\omega +i\eta +E_{\mathbf{k}}^{-}}\right) ;  \label{Gsym} \\
Z_{\mathbf{k}}^{\pm } &=&\epsilon _{\mathbf{k}}/\sqrt{\Delta _{pg}^{2}+\left
\vert \epsilon _{\mathbf{k}}\right \vert ^{2}}\pm 1\text{,}  \notag
\end{eqnarray}%
where $\epsilon _{\mathbf{k}}$ was defined in Eq.(\ref{epsilondef}) and $%
\eta $ is the damping parameter. The dispersion relation in the nonmagnetic
basis takes a form%
\begin{equation}
E_{\mathbf{k}}^{\pm }=\epsilon _{\mathbf{k}}^{\prime }-\mu +U_{r}\frac{n}{2}%
\pm \sqrt{\Delta _{pg}^{2}+\left \vert \epsilon _{\mathbf{k}}\right \vert
^{2}}\text{,}  \label{Enonmag}
\end{equation}%
where $\epsilon _{\mathbf{k}}$ was defined in Eq.(\ref{epsilondef}). This is
quite similar to one obtained in the slave boson approach to the t-J\cite%
{slave} and RVB\cite{ZhangRVB} approaches.

\subsection{EPI in the magnetic Brillouin zone}

The connection between the electron - electron attraction due to phonons
given in Eq.(\ref{Aee}) in the usual "paramagnetic" basis, that is full BZ
(marked by $\overline{v}_{k_{x},k_{y}}$ here) in the underdoped cases should
be represented as a matrix elements in the sublattice space defined on a
smaller magnetic BZ. The matrix,

\begin{equation}
v_{k_{1}k_{2}}^{ph}=\frac{1}{2}%
\begin{pmatrix}
\overline{v}_{k_{1},k_{2}-k_{1}}+\overline{v}_{k_{1}+\pi ,k_{2}-k_{1}+\pi }
& \left( \overline{v}_{k_{1},k_{2}-k_{1}}-\overline{v}_{k_{1}+\pi
,k_{2}-k_{1}+\pi }\right) \exp \left[ -iak_{1}\right] \\ 
\left( \overline{v}_{k_{1},k_{2}-k_{1}}-\overline{v}_{k_{1}+\pi
,k_{2}-k_{1}+\pi }\right) \exp \left[ iak_{1}\right] & \overline{v}%
_{k_{1},k_{2}-k_{1}}+\overline{v}_{k_{1}+\pi ,k_{2}-k_{1}+\pi }%
\end{pmatrix}%
\text{,}  \label{B_vph1}
\end{equation}%
was used to calculate the phonon effects in both normal and superconducting
state.

\subsection{Susceptibility}

The susceptibility matrix that enters the effective electron - electron
interaction strength due to (the Hubbard repulsion induced) correlations is
calculated in the post-Gaussian approximation as Lindhard type diagrams
given in Fig.11. They are similar to the paramagnetic case\cite{Kohn}\cite%
{Maier20}. The propagators of the diagrams however, Eq.(\ref{AFcorr}), are
defined on magnetic BZ and have two sublattice indices. The spin singlet
pairing contribution to elements comes from the left and center diagrams:

\begin{eqnarray}
\chi _{m,\mathbf{q}}^{II} &=&\frac{T}{NN^{\prime }}\sum \nolimits_{n\mathbf{p%
}}\left( -G_{m+n,\mathbf{q+p}}^{\downarrow II}G_{n\mathbf{p}}^{\downarrow
II}+G_{m+n,\mathbf{q+p}}^{\downarrow II}G_{n\mathbf{p}}^{\uparrow II}\right)
;  \label{B_chai} \\
\chi _{l,q}^{AB} &=&-\frac{T}{NN^{\prime }}\sum \nolimits_{n\mathbf{p}%
}\left( -G_{m+n,\mathbf{q+p}}^{\downarrow AB}G_{n\mathbf{p}}^{\downarrow
BA}+G_{m+n,\mathbf{q+p}}^{\downarrow AB}G_{n\mathbf{p}}^{\uparrow BA}\right)
;  \notag \\
\chi _{l,\mathbf{q}}^{BA} &=&\chi _{-l,-\mathbf{q}}^{AB\ast },  \notag
\end{eqnarray}%
where $N^{\prime }=N/2$ and $I=A,B$. The third diagram vanishes.

Summing up over integers $n$, one obtains

\begin{eqnarray}
\chi _{m\mathbf{q}}^{AA} &=&P_{m\mathbf{q}}-Q_{m\mathbf{q}};
\label{B_chaires} \\
\chi _{m\mathbf{q}}^{AB} &=&R_{m\mathbf{q}};  \notag \\
\chi _{m\mathbf{q}}^{BB} &=&-P_{m\mathbf{q}}-Q_{m\mathbf{q}}\text{,}  \notag
\end{eqnarray}%
where%
\begin{eqnarray}
P_{m\mathbf{q}} &=&\frac{1}{2NN^{\prime }}\sum \nolimits_{\mathbf{p}}\left(
L_{m}\left[ E_{\mathbf{p}}^{-},E_{\mathbf{q+p}}^{-}\right] +L_{m}\left[ E_{%
\mathbf{p}}^{-},E_{\mathbf{q+p}}^{+}\right] +L_{m}\left[ E_{\mathbf{p}%
}^{+},E_{\mathbf{q+p}}^{-}\right] +L_{m}\left[ E_{\mathbf{p}}^{+},E_{\mathbf{%
q+p}}^{+}\right] \right) ;  \label{B_PQR} \\
Q_{m\mathbf{q}} &=&\frac{\Delta _{pg}}{NN^{\prime }}\sum \nolimits_{\mathbf{p%
}}\frac{1}{x_{\mathbf{p}}}\left( L_{m}\left[ E_{\mathbf{p}}^{+},E_{\mathbf{%
q+p}}^{-}\right] -L_{m}\left[ E_{\mathbf{p}}^{-},E_{\mathbf{q+p}}^{+}\right]
+L_{m}\left[ E_{\mathbf{p}}^{+},E_{\mathbf{q+p}}^{+}\right] -L_{m}\left[ E_{%
\mathbf{p}}^{-},E_{\mathbf{q+p}}^{-}\right] \right) ;  \notag \\
R_{m\mathbf{q}} &=&\frac{1}{2NN^{\prime }}\sum \nolimits_{\mathbf{p}}\frac{%
h_{\mathbf{q+p}}^{\ast }h_{\mathbf{p}}}{x_{\mathbf{q+p}}x_{\mathbf{p}}}%
\left( L_{m}\left[ E_{\mathbf{p}}^{-},E_{\mathbf{q+p}}^{+}\right] +L_{m}%
\left[ E_{\mathbf{p}}^{+},E_{\mathbf{q+p}}^{-}\right] -L_{m}\left[ E_{%
\mathbf{p}}^{+},E_{\mathbf{q+p}}^{+}\right] -L_{m}\left[ E_{\mathbf{p}%
}^{-},E_{\mathbf{q+p}}^{-}\right] \right) \text{.}  \notag
\end{eqnarray}%
Here 
\begin{equation}
L_{m}\left[ E_{1},E_{2}\right] =\frac{f_{F}\left[ E_{1}\right] -f_{F}\left[
E_{2}\right] }{2i\pi Tm+E_{1}-E_{2}}\text{,}  \label{B_Jdef}
\end{equation}%
$\Delta _{pg}$ is the pseudogap energy, $x_{\mathbf{p}}$ is defined in Eq.(%
\ref{xdef}), $E_{\mathbf{p}}^{\pm }$ in \ Eq.(\ref{Edef}) and $h_{\mathbf{p}%
} $ in Eq.(\ref{hdef}).

\begin{figure}[h]
\centering \includegraphics[width=16cm]{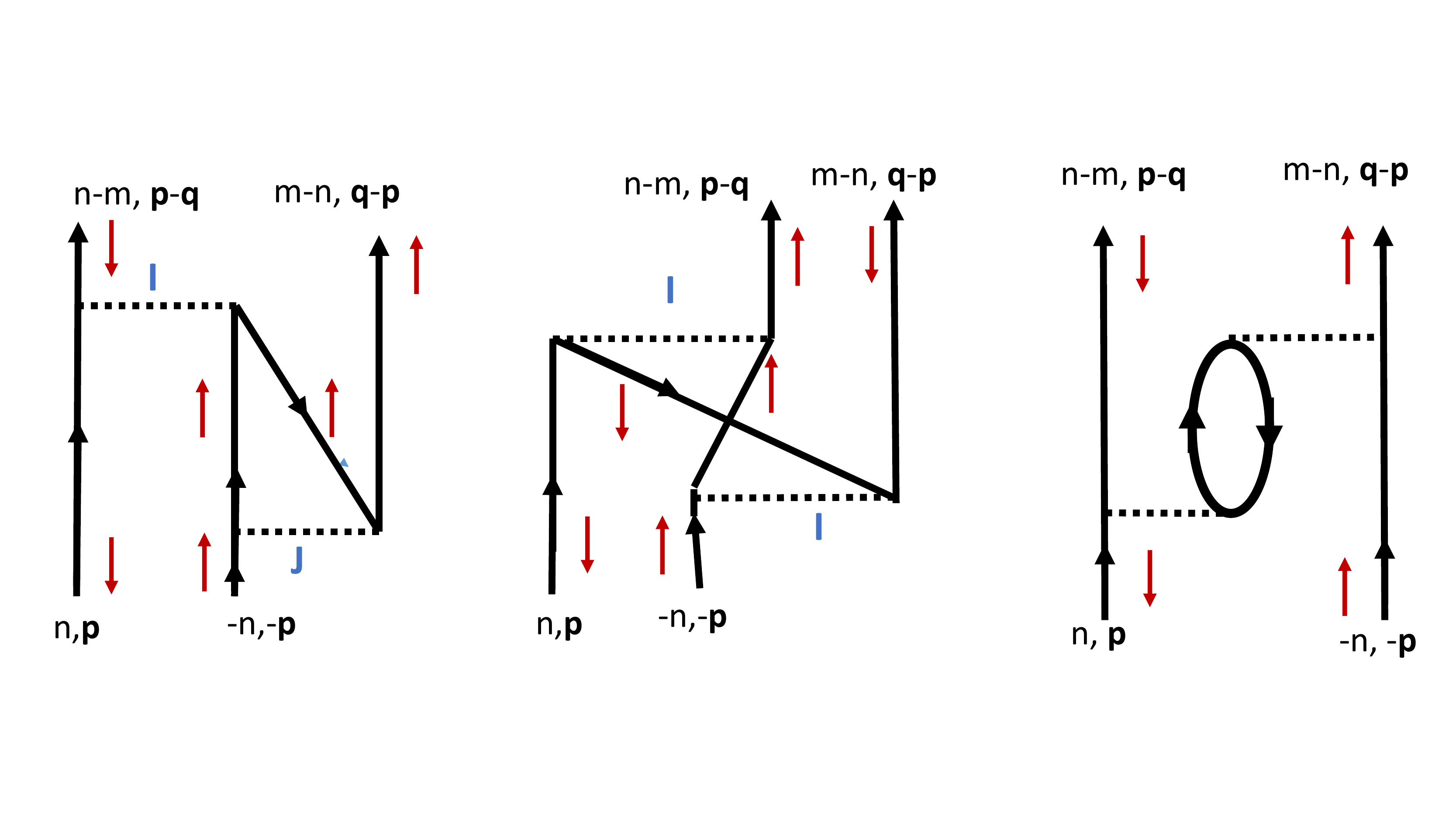}
\caption{Three second order diagrams determining the effective electron -
electron interaction due to spin fluctuations. Botth spin and sublattice
indices are indicated. While the diagram on left and center give nonzero
contributions of Eq.(\protect \ref{B_chai}), the third vanishes due to
conflict in assigning spin indices to propagators in the loop.}
\end{figure}

\subsection{Kink in dispersion relation (overdoped)}

The calculation is similar to that for the simpler paramagnetic case. The
result is:

\begin{eqnarray}
\frac{d}{d\omega }I_{\omega ,\mathbf{p}} &=&\frac{Z^{+}}{2}\left( \frac{f_{B}%
\left[ \Omega \right] -f_{F}\left[ -E_{\mathbf{k+l}}^{+}\right] +1}{\left(
\omega +i\eta +\Omega -E_{\mathbf{k+l}}^{+}\right) ^{2}}+\frac{f_{B}\left[
\Omega \right] +f_{F}\left[ -E_{\mathbf{k+l}}^{+}\right] }{\left( \omega
+i\eta -\Omega -E_{\mathbf{k+l}}^{+}\right) ^{2}}\right)  \notag \\
&&+\frac{Z^{-}}{2}\left \{ E^{+}\rightarrow E^{-}\right \} \text{,}
\end{eqnarray}%
where the energies $E^{\pm }$ and $Z^{\pm }$ were defined in Eq.(\ref{Gsym},%
\ref{Enonmag}).

\section{C. Gap equation in underdoped system}

\subsection{Derivation}

We derive the Gorkov's equations within the functional integral approach\cite%
{NO} starting from the effective electron action for grassmanian fields $%
\psi _{\mathbf{k},n}^{\ast \sigma }$ and $\psi _{\mathbf{k},n}^{\sigma }$.
To simplify the presentation it is useful to lump the quasi - momentum and
the Matsubara frequency into a single subscript, $\left \{
n,k_{1},k_{2}\right \} \rightarrow \alpha ,$and the spin and sublattice into
the four component spinor $\left \{ \sigma ,I\right \} \rightarrow a$. The
action of Eq.(\ref{actionAF}) takes a standard multicomponent four - Fermi
form studied for example in ref.\cite{Rosen19}:

\begin{equation}
\mathcal{A}\left[ \psi \right] =\psi _{\alpha }^{\ast a}T_{\alpha }^{ab}\psi
_{\alpha }^{b}+\frac{1}{2}\psi _{\beta }^{\ast a}\psi _{\chi +\beta
}^{a}v_{-\chi }^{ab}\psi _{\gamma }^{\ast b}\psi _{\gamma -\chi }^{b}\text{.}
\label{C_action}
\end{equation}%
The hopping $4\times 4$ matrix (inverse GF) for $a=\left \{ \sigma
,I\right
\} $, $b=\left \{ \rho ,J\right \} $ in the following form,

\begin{equation}
T_{\left \{ n,k_{1},k_{2}\right \} }^{ab}=%
\begin{array}{cc}
\delta ^{\sigma \rho }\left( -i\omega _{n}+\varepsilon _{k}^{\prime }-\mu +%
\frac{U_{r}n}{2}\right) +\sigma _{z}^{\sigma \rho }\Delta _{pg} & -\delta
^{\sigma \rho }h_{k}^{\ast } \\ 
-\delta ^{\sigma \rho }h_{k} & \delta ^{\sigma \rho }\left( -i\omega
_{n}+\varepsilon _{k}^{\prime }-\mu +\frac{U_{r}n}{2}\right) -\sigma
_{z}^{\sigma \rho }\Delta _{pg}%
\end{array}%
\text{,}  \label{C_hopping}
\end{equation}%
with $I$ and $J$ being the row and the column indices.

Gorkov equations in matrix form are:%
\begin{eqnarray}
-G_{\alpha }T_{\alpha }-F_{\alpha }\Delta _{\alpha }^{\ast t} &=&I;
\label{C_Gorkov} \\
G_{\alpha }\Delta _{\alpha }-F_{\alpha }T_{-\alpha }^{t} &=&0\text{,}  \notag
\end{eqnarray}%
where $F_{\alpha }$ is the anomalous GF and the matrix gap function is
defined $\left[ \Delta _{\alpha }\right] $ in components as (see Fig.12))%
\begin{equation}
\Delta _{\alpha }^{bc}=\sum \nolimits_{\chi }v_{\alpha -\chi }^{bc}F_{\chi
}^{bc}\text{.}  \label{C_delta}
\end{equation}%
The corresponding gap equation is%
\begin{equation}
\Delta _{\alpha }^{bc}=-\sum \nolimits_{\chi }v_{\alpha -\chi }^{bc}\left[
\left( T_{-\chi }^{t}\left[ \Delta _{\chi }\right] ^{-1}T_{\chi }+\Delta
_{\chi }^{\dagger }\right) ^{-1}\right] ^{bc}\text{.}  \label{C_gapeq}
\end{equation}%
The singlet Ansatz Eq.(\ref{delAF}) leads to Eq.(\ref{GapeqAF}).\newpage

\begin{figure}[h]
\centering \includegraphics[width=10cm]{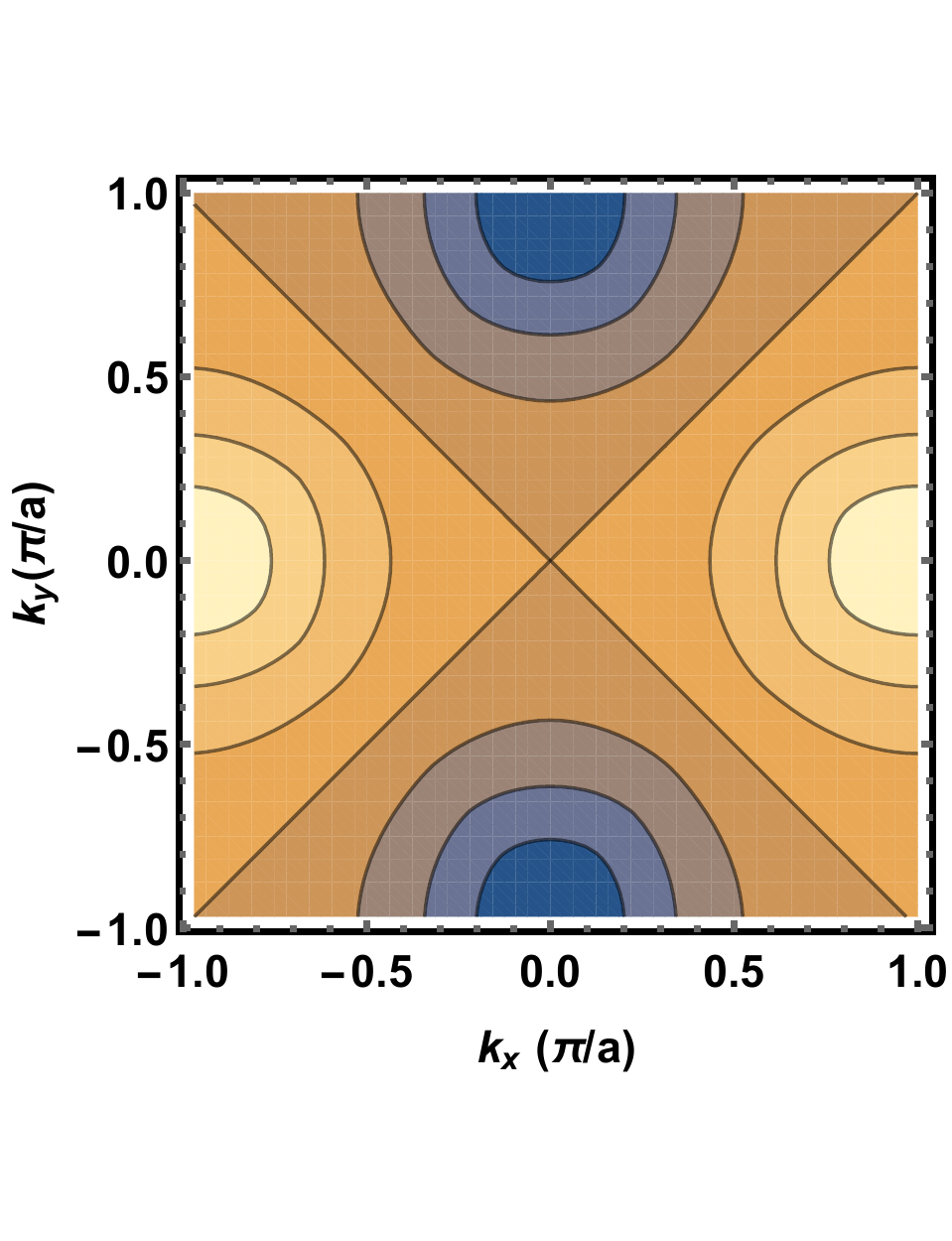}
\caption{The d - wave solution of the gap equation for optimal doping, $%
x=0.166$ at $50K$.}
\end{figure}

\begin{figure}[h]
\centering \includegraphics[width=14cm]{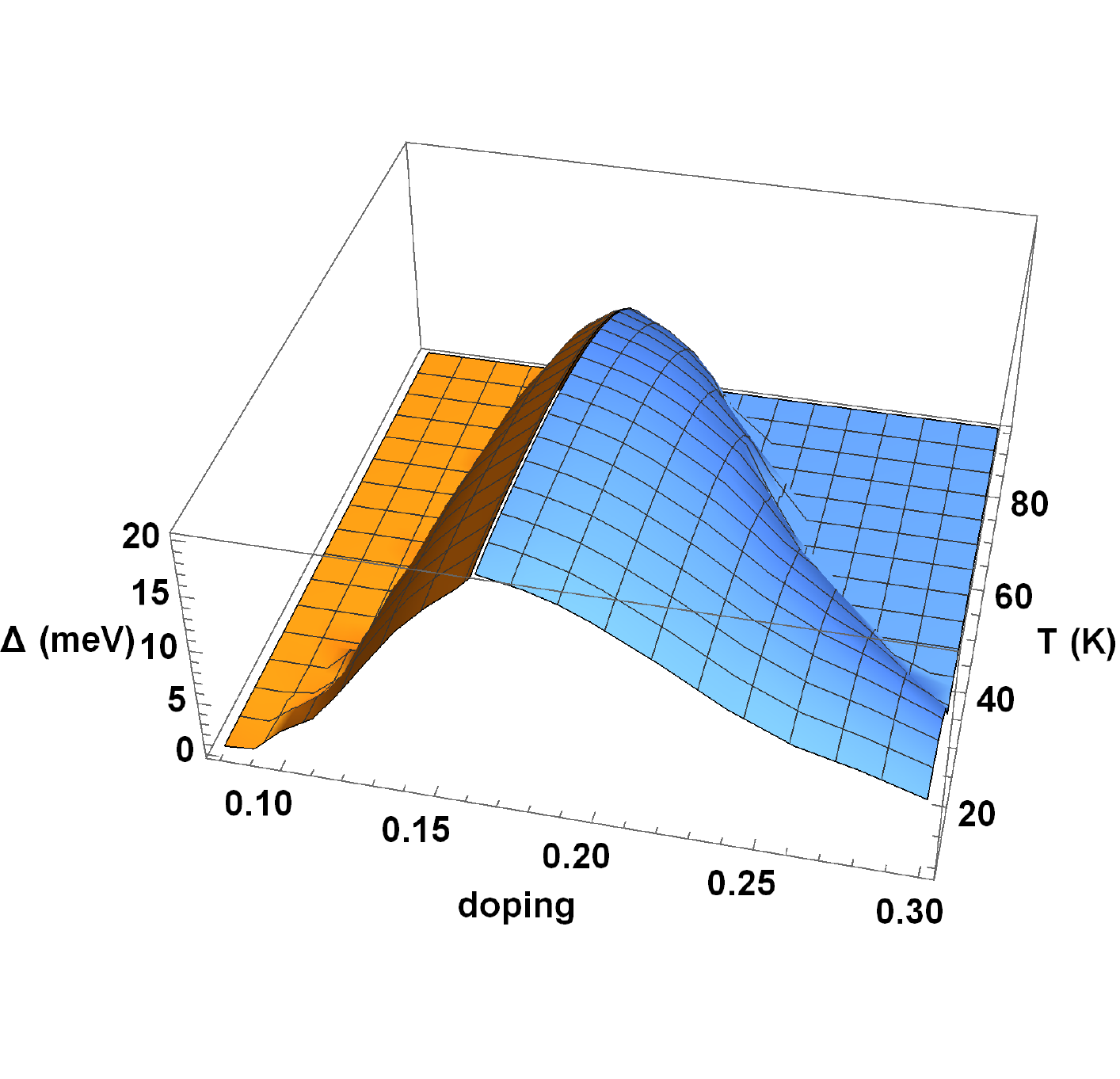}
\caption{ Superconducting (maximal) d - wave Matsubara gap as function of
dopings and temperatures. Underdoped parts are in brown, while the overdoped
in blue.}
\end{figure}

\subsection{The d -wave gap}

The blue part of the surface corresponds to $x\geq x^{opt}$. The line of
vanishing gap determines the critical temperature values on the phase
diagram in Fig. 4 (red squares). In the optimal and overdoped domains it
agrees well with the parabolic experimental dependence (dashed curve) taken
from ref.\cite{accurate}. If one neglects the magnon contribution, namely
takes $v=v^{ph}$, the temperatures are lower by 15-20\% (red circles).

One observes that the decrease of $T_{c}$ is rather slow (linear) at large
doping compared to the experiment. When doping becomes of order 30\% it is
expected to significantly impacts the effective mesoscopic lattice model
parameters ($\mu ,U,t,t^{\prime }$). In underdoped cases the pseudogap
should be taken into account. The results are the yellow part of the surface
in Fig.13 for the gap and critical temperatures shown on the left hand side
of the phase diagram, Fig.3. The maximum gap as function of doping and
temperature is given in Fig.13 (the yellow part of the surface).

\end{document}